\documentclass[a4paper, 12pt]{article}

\usepackage[english]{babel}
\usepackage[sort&compress]{natbib}
\bibpunct{(}{)}{;}{a}{}{,} 

\usepackage{amsthm, amsmath, amssymb, mathrsfs, multirow, url, subfigure}
\usepackage{graphicx} 
\usepackage{ifthen} 
\usepackage{amsfonts}
\usepackage[usenames]{color}
\usepackage{fullpage}
\usepackage{booktabs}
\usepackage[ruled,vlined]{algorithm2e}

\theoremstyle{plain}

\theoremstyle{definition}

\theoremstyle{remark}

\newcommand{\E}{\mathsf{E}}

\newcommand{\nm}{{\sf N}}

\newcommand{\gam}{{\sf Gamma}}
\newcommand{\stt}{{\sf t}}

\newcommand{\logn}{{\sf logN}}
\newcommand{\Chisq}{{\sf ChiSq}}
\newcommand{\lap}{{\sf Lap}}

\newcommand{\RR}{\mathbb{R}}

\newcommand{\YY}{\mathbb{Y}}

\newcommand{\C}{\mathcal{C}}

\renewcommand{\S}{\mathcal{S}}

\newcommand{\eps}{\varepsilon}

\newcommand{\model}{\mathscr{P}}

\usepackage{stackrel}

\title{Calibrating generalized predictive distributions}
\author{Pei-Shien Wu\footnote{Department of Statistics, North Carolina State University; {\tt pwu9@ncsu.edu}, {\tt rgmarti3@ncsu.edu}} \quad and \quad Ryan Martin$^*$}
\date{\today}

\begin{document}

\maketitle 

\begin{abstract}    
In prediction problems, it is common to model the data-generating process and then use a model-based procedure, such as a Bayesian predictive distribution, to quantify uncertainty about the next observation.  However, if the posited model is misspecified, then its predictions may not be calibrated---that is, the predictive distribution's quantiles may not be nominal frequentist prediction upper limits, even asymptotically.  Rather than abandoning the comfort of a model-based formulation for a more complicated non-model-based approach, here we propose a strategy in which the data itself helps determine if the assumed model-based solution should be adjusted to account for model misspecification.  This is achieved through a generalized Bayes formulation where a learning rate parameter is tuned, via the proposed {\em generalized predictive calibration} (GPrC) algorithm, to make the predictive distribution calibrated, even under model misspecification.  Extensive numerical experiments are presented, under a variety of settings, demonstrating the proposed GPrC algorithm's validity, efficiency, and robustness.


\smallskip

\end{abstract}

\section{Introduction}
\label{S:intro}  

Prediction of future observations is a common goal in applications and is a fundamental problem in statistics and machine learning.  Motivated by the ``all models are wrong'' part of the famous quote attributed to George Box, many researchers advocate for model-free methods that can make point predictions in complex problems without the risk of model misspecification biases.  However, without specification of a statistical model, quantification of prediction uncertainty---via valid prediction intervals or, in our present case, valid predictive distributions---can be a challenge.  Therefore, motivated by the practical need for prediction uncertainty quantification \citep[e.g.,][]{patel1989prediction, tian2020methods} and the ``but some models are useful'' part of the famous quote, many researchers advocate for the use of (carefully specified) statistical models and the associated model-based methods.  But despite the data analyst's best efforts to specify a sound model, misspecification biases are unavoidable.  Is it possible for a model-based method to have a built-in correction that will adjust, and in a data-driven way, the prediction intervals/predictive distribution to account for potential model misspecification biases?  If so, then this would provide a sort of ``best of both worlds,'' that is, data analysts can work in a familiar model-based framework that readily provides prediction uncertainty quantification with the added comfort that their predictive inference will, in a certain sense, remain valid even when the model is misspecified.  Development of this built-in correction of model-based methods---in particular, of Bayesian predictive distributions---to achieve the aforementioned ``best of both worlds'' is the goal of the present paper.  

Our motivation for this work was thinking about insurance applications where the insurance company's goal is to predict the loss (total claim amount) in the next time unit based on observed losses in previous time units, often treated as independent and identically distributed (iid).  In these applications, it is common for the loss distribution to be heavy-tailed, and a host of different models, and the associated model-based (likelihood and Bayesian) methods have been developed specifically for this \citep[e.g.,][]{bk.2016, frees.book, klugman, cooray2018}.  Of course, even a carefully-specified heavy-tailed model could be misspecified, so some authors have proposed more flexible---and more complicated---nonparametric methods that let the data decide the distributional form \citep[e.g.,][and the references therein]{hong2018dirichlet, hong.martin.recursive}.  At the end of the day, however, practitioners want their models/methods to be {\em exactly} as complicated as necessary.  Could, for example, the tail of a simple, thin-tailed predictive distribution be fattened if necessary, in a data-driven way, to adjust for possibly heavy-tailed data?  Similar questions arise in other applications, such as spatial data analysis, especially when interest is in spatially and temporally dependent extremes; see Section~\ref{S:result.spatial} for details and references.

Our starting point is a Bayesian formulation wherein, if the model is correctly specified and satisfies certain regularity conditions, then the posterior distribution will concentrate around the true parameter value asymptotically and, consequently, the corresponding Bayesian predictive distribution will merge with the true data-generating distribution.  In this case, we say that the Bayesian predictive distribution is valid or {\em calibrated}, at least asymptotically, in the sense that the upper-$\alpha$ quantile of the predictive distribution is an approximate $100(1-\alpha)$\% upper prediction limit.  When the Bayesian model is misspecified, however, the situation is not so straightforward \citep[e.g.,][]{bunke1998asymptotic,kleijn2006misspecification, kleijn2012bernstein,grunwald.ommen.scaling, rvr.sriram.martin}.  To start, there is no ``true parameter value'' so the posterior distribution cannot possibly concentrate there.  Moreover, since the model is wrong, it is not possible for the Bayesian predictive distribution to merge with the true distribution, so the latter generally will not be calibrated in the sense defined above.  To help overcome this model misspecification bias, a number of authors have recently been using a so-called {\em generalized posterior}, which amounts to using a power likelihood in the Bayesian update \citep[e.g.,][]{miller.dunson.power, bhat.pati.yang.fractional,thomas2019diagnosing}.  This power, denoted by $\eta$, is called the {\em learning rate} and must be chosen using the data.  But note that $\eta$ is not a model parameter with a true value that can be ``learned'' in any meaningful way, e.g., using a prior and Bayesian updates.  Instead, $\eta$ is a tuning parameter and must be chosen to achieve some desirable operating characteristic.  A number of learning rate selection procedures have been developed in the literature recently, including \citet{holmes2017assigning}, \citet{lyddon2019general},
\citet{grunwald.ommen.scaling}, and 
\citet{syring2019calibrating}; see \citet{wu2020comparison} for a comparison.  For sure, whatever operating characteristics are achieved for the generalized posterior through these choices of $\eta$ do not automatically carry over to the corresponding predictive distribution, so it is necessary to reconsider things when prediction is the goal.  Here we make two contributions: first, we define a suitable $\eta$-{\em generalized predictive distribution}---see \eqref{eq:gpred}---for which calibration is possible; second, inspired by \citet{syring2019calibrating}, we develop a learning rate selection procedure designed so that calibration in the above sense is achieved.

The remainder of this paper is organized as follows. First, in Section~\ref{S:background}, we provide some background on generalized posterior distributions and the recently developed methods for learning rate selection. In Section~\ref{S:predictive}, we define our generalized predictive distribution, introduce our predictive distribution calibration method, and formulate the corresponding {\em generalized predictive calibration algorithm}, or GPrC for short. The performance of our GPrC method is demonstrated in a variety of examples with different kinds of model misspecification. First, some relatively simple illustrations using independent and identically distributed (iid) models are presented in Section~\ref{S:result.simple} to highlight the GPrC algorithm's ability to adjust a relatively simple, thin-tailed model to accommodate heavy-tailed data. Then, in Section~\ref{S:more}, we extend our generalized predictive distribution and GPrC algorithm formulation to adjust for model misspecification in more complex, non-iid models used for analyzing time series and spatial data. In particular, in Section~\ref{S:result.spatial}, we consider prediction of the response evaluated at a new spatial location, where we posit a simple Gaussian process model and, using the GPrC algorithm, are able to achieve valid predictions, even in the extreme tails and under very non-Gaussian data-generating processes. Finally, some concluding remarks are given in Section~\ref{S:discuss}.

\section{Background}
\label{S:background}

\subsection{Generalized posterior distributions}

For simplicity, suppose that we have iid data $Y_1,\ldots,Y_n$, taking values in a space $\YY$, with common marginal distribution $P^\star$; later we will consider more general cases with dependence and/or covariates.  The primary goal is to predict the next observation, $Y_{n+1}$. For the purpose of analyzing data and ultimately making predictions, it is common to introduce a statistical model, $\model = \{P_\theta: \theta \in \Theta\}$, a family of distributions on $\YY$, indexed by a parameter $\theta$ in the parameter space $\Theta$; let $p_\theta$ denote the density of $P_\theta$ with respect to some dominating $\sigma$-finite measure on $\YY$, such as Lebesgue or counting measure.  The  prediction problem basically amounts to learning $P^\star$ from data, but the introduction of a model shifts the focus to the ``true value'' $\theta^\star$ of $\theta$.  Then the likelihood function $L_n$---which, in this iid setting, is $L_n(\theta) = \prod_{i=1}^n p_\theta(Y_i)$---plays an important role in any model-based approach.  One such approach is Bayesian, which proceeds by introducing a prior distribution $\Pi$ on $\Theta$ that quantifies the {\em a priori} uncertainty about the unknown value of $\theta$.  This prior is then combined with the likelihood, via Bayes's formula, to get a posterior distribution 
\[ \Pi_n(d\theta) \propto L_n(\theta) \, \Pi(d\theta), \quad \theta \in \Theta. \]
If the model is sufficiently regular and correctly specified, i.e., if $P^\star\in\model$, then the Bayesian posterior distribution has nice asymptotic properties. That is, the Bernstein--von Mises theorem \citep[e.g.,][Chapter 10.2]{van2000asymptotic} implies that $\Pi_n$ is asymptotically normal with mean equal to $\hat{\theta}_n$, the maximum likelihood estimator, and covariance matrix proportional to the inverse of the Fisher information matrix. Since $\hat\theta_n$ is typically consistent, this implies that $\Pi_n$ will be centered around $\theta^\star$ and that the corresponding posterior credible sets are asymptotically correct confidence sets.  

In practice, however, one can never be sure that the model is correctly specified, so it is practically relevant to consider what happens if the model is misspecified, i.e., if $P^\star \not\in \model$.  An immediate consequence is that there is no ``true'' $\theta^\star$, so it is not clear what the Bayesian posterior distribution $\Pi_n$ might be learning about.  However, under certain regularity conditions, a misspecified version of the aforementioned Bernstein--von Mises theorem holds \citep{kleijn2012bernstein}, which states that $\Pi_n$ is still asymptotically normal, centered at $\hat\theta_n$, with covariance matrix $n^{-1} V$; the specific form of $V$ is known---it depends on the form of $P_\theta$---but the expression is not needed for what follows.  Moreover, $\hat\theta_n$ is a consistent estimator of the Kullback--Leibler minimizer 
\[ \theta^\dagger = \arg\min_{\theta \in \Theta} \int \log(dP^\star / dP_\theta) \, dP^\star, \]
but its asymptotic covariance matrix is generally different from $n^{-1} V$.  Therefore, while inference on $\theta^\dagger$ would be meaningful---since $P_{\theta^\dagger}$ is the ``best approximation'' of $P^\star$ in $\model$---the covariance matrix mismatch implies that posterior credible sets derived from $\Pi_n$ could have arbitrarily low coverage probability.  

Is there anything that can be done to correct for the effects of model misspecification, short of starting over from scratch with a different $\model$?  An idea that has gained some traction in the recent literature is the use of a so-called {\em generalized posterior}, which uses Bayes's formula but with a power likelihood:
\begin{equation}
\label{eq:gbayes}
\Pi_n^{(\eta)}(d\theta) \propto L_n^\eta(\theta) \, \Pi(d\theta), \quad \theta \in \Theta, \quad \eta > 0.
\end{equation}
The power $\eta$ is commonly referred to as the {\em learning rate}.  Even in correctly specified models, there is an advantage to working with the generalized posterior, since, for any $\eta < 1$, the asymptotic concentration properties of $\Pi_n^{(\eta)}$ can be established without entropy conditions \citep[e.g.,][]{walker.hjort.2001, zhang2006a, grunwald.mehta.rates}; see \citet{martin.walker.deb} for a different use of power likelihood.  When the model is potentially misspecified, the likelihood function $L_n$ does not hold the same stature as it does in well-specified cases, so its role in the Bayesian update is less clear.  But if one interprets the Kullback--Leibler minimizer $\theta^\dagger$ as the quantity of interest, the problem then can be viewed as one of ``risk-minimization,'' with $\ell_\theta(y) = -\log p_\theta(y)$ as the loss.  From this perspective, the generalized posterior \eqref{eq:gbayes} is also a so-called {\em Gibbs posterior} \citep[e.g.,][]{jiang2008gibbs, gibbs.general}, which has been shown to be the proper coherent updating of prior information under misspecification; see \citet{walker2013} and \citet{bissiri.holmes.walker.2016}.  Therefore, the learning rate is a fundamental part of Bayesian inference in misspecified models.  

Roughly speaking, $\eta$ controls the spread of the generalized posterior.  That is, if the learning rate is too large, then the generalized posterior will be tightly concentrated around $\hat\theta_n$; conversely, if the learning rate is small, then the prior gets more weight and generally this will make the generalized posterior less concentrated around $\hat\theta_n$.  So, clearly, the choice of $\eta$ affects the practical, finite-sample properties of the generalized posterior.  This begs the question: how to choose $\eta$?


\subsection{Learning rate selection methods}
\label{SS:eta}

A number of different data-driven methods for selecting the learning rate $\eta$ have been proposed in the recent literature.  Here we give just a brief summary of these; a more thorough review can be found in \citet{wu2020comparison}.

\begin{itemize}

\item Motivated by the developments in \citet{bissiri.holmes.walker.2016}, {\em \citet{holmes2017assigning}} proposed to choose $\eta$ by matching the prior-to-posterior expected information gain between ordinary Bayesian and generalized posterior.  By measuring this gain in terms of the Fisher divergence, they are able to solve this equation for $\eta$.  This expression involves certain unknowns, but they provide simple, Monte Carlo estimates for these unknowns, leading to a data-driven learning rate $\hat\eta$. 

\vspace{-2mm}

\item Motivated by the prequential view of Bayesian model selection in \citet{dawid1984present}, {\em \citet{grunwald.ommen.scaling}} proposed the {\em SafeBayes} algorithm that chooses $\eta$ by minimizing a minor variation on the cumulative log-loss, i.e., 
\[ \hat\eta = \arg\min_\eta \sum_{i=1}^n \int \{-\log p_\theta(Y_i)\} \, \Pi_{i-1}^{(\eta)}(d\theta). \]


\vspace{-2mm}

\item 
The learning rate selection method is viewed as the covariance mismatch problem in \cite{lyddon2019general}. Motivated by the asymptotic properties of weighted likelihood bootstrap method \citep[e.g.,][]{newton1994approximate} under well-specified model, they developed a loss-likelihood bootstrap suitable for cases involving misspecified models. By the asymptotic distribution properties of the loss-likelihood bootstrap samples, \citet{lyddon2019general} proposed to match its covariance matrix with the asymptotic covariance matrix of the generalized Bayesian posterior. 

\vspace{-2mm}

\item {\em \citet{syring2019calibrating}} aimed specifically to choose the learning rate $\eta$ to partially correct for the aforementioned covariance matrix mismatch, so that  $\eta$-generalized posterior credible sets were calibrated in the sense that the frequentist coverage probability is approximately the nominal level.  They achieved this through a {\em generalized posterior calibration} algorithm, or GPC, which approximates the coverage probability function using bootstrap and then uses a version of stochastic approximation (see Section~\ref{SS:gprc} below) to match that coverage to the nominal level.  While both GPC and Lyddon et al.~use bootstrap, the methods are, in fact, quite different; in particular, GPC more directly targets calibration of posterior credible regions. 


\end{itemize}
The overall conclusion drawn in \citet{wu2020comparison} is that only the GPC algorithm provides satisfactory calibration of the generalized posterior credible sets in general.  This is not surprising, given that the other methods are designed to achieve other properties.  However, if the generalized posterior's purpose is to provide valid, data-driven uncertainty quantification about the unknowns, then this kind of calibration is essential. Therefore, in what follows, we aim to develop a prediction-focused analogue of calibration and the GPC algorithm developed in \citet{syring2019calibrating}.


\section{Generalized predictive distributions}
\label{S:predictive}

\subsection{Objective: calibration}

Suppose we have data $Y^n = (Y_1,\ldots,Y_n)$ generated from a true distribution $P^\star$, and the primary goal is prediction of the next observation, $Y_{n+1}$.  While prediction is the primary goal, rather than inference, common practice is to introduce a statistical, $\model = \{P_\theta: \theta \in \Theta\}$.  From here, as described in Section~\ref{SS:def} below, we will construct a model-based predictive distribution for $Y_{n+1}$, depending of course on the observed data $Y^n$. Here we explain what operating characteristics we hope this predictive distribution to have, whether the model is correctly specified or not.  

For $\alpha \in (0,1)$, let $Q_\alpha(y^n)$ denote the upper-$\alpha$ quantile of this predictive distribution, which is a function of $y^n$.  Then, following \citet{dawid1984present}, \citet{grunwald.safe}, and \citet{tian2020methods}, we say that the predictive distribution is {\em calibrated} (at level $\alpha$) if 
\begin{equation}
\label{eq:calibration}
 P^\star\{ Q_\alpha(Y^n)\geq Y_{n+1}\} \geq 1-\alpha, 
 \end{equation}
that is, if the set to which the predictive distribution assigns probability $1-\alpha$ has frequentist coverage probability at least $1-\alpha$. Obviously, if $P^\star$ were known, then we could take the upper-$\alpha$ quantile, $Q_\alpha^\star(y^n)$, of the corresponding conditional distribution $P^\star(\cdot \mid y^n)$, and it would be calibrated in the sense of \eqref{eq:calibration}; in the iid case, this would simplify because $Q_\alpha^\star$ would be a constant, independent of $y^n$.  

In practice, however, the true distribution $P^\star$ is unknown, so calibration is non-trivial. In what follows, we develop a framework in which a model-based predictive distribution can be tuned to accommodate potential model misspecification in such a way that calibration is achieved, at least approximately. This will rely on ideas that parallel those described above for generalized posteriors and learning rate selection. At least at a high level, our setup closely resembles that in \citet{grunwald.safe} in the following sense: we are starting with a simple, pragmatic Bayesian model, which we readily acknowledge may not be correctly specified, and then develop a data-driven adjustment to our pragmatic model's predictive distribution so that it is reliable---or, in Gr\"unwald's words, ``safe''---at least in the sense that \eqref{eq:calibration} is satisfied.


\subsection{Definition}
\label{SS:def}

There are a number of ways one might consider defining a generalized predictive distribution. A relative general umbrella that these different ideas fall under is to define a predictive distribution indexed by a trio of positive scalars $(a,b,c)$ as follows:
\[ f_n^{(a,b,c)}(y) \propto \Bigl\{ \int p_\theta^a(y) \, \Pi_n^{(b)}(d\theta) \Bigr\}^{1/c}. \]
Here, $p_\theta^a$ is the model density to power $a > 0$, $\Pi_n^{(b)}$ is the generalized posterior in \eqref{eq:gbayes} with learning rate $b > 0$, and the proportionality constant is determined by integrating the right-hand side with respect to $y$. Then different ideas for constructing a generalized predictive distribution correspond to different configurations of $(a,b,c)$.  
\begin{enumerate}
\item An ordinary Bayes predictive distribution corresponds to $a=b=c=1$.

\vspace{-2mm}

\item A natural generalization of the Bayesian predictive distribution is to take $a=c=1$ and let $b=\eta$ be a free learning rate parameter to be chosen using, e.g., one of the procedures described in Section~\ref{SS:eta}.  

\vspace{-2mm}

\item Given that the generalized posterior has the model density $p_\theta$ to a power $\eta$, it also makes sense to consider the same power on the model density when forming the predictive distribution. This corresponds to $c=1$ and $a=b=\eta$, with $\eta$ a tuning parameter to be selected.

\vspace{-2mm}

\item \citet{corcuera1999generalized, corcuera1999relationship} considered the case where $b=1$ and $a=c=\frac12(1-\beta)$, where $\beta$ indexes the user's choice of divergence measure. With this choice of $(a,b,c)$, the corresponding $f_n^{(a,b,c)}$ is the Bayes estimator, i.e., posterior risk minimizer, under the so-called ``$\beta$-divergence'' loss, where $\beta=-1$ corresponds to Kullback--Leibler divergence, $\beta=0$ corresponds to Hellinger distance, etc. The specific form of this divergence is not important for us here. A further generalization was presented in \citet{zhang2017information}, which basically takes $b=\frac12(1+\beta)$; we say ``basically'' because they actually take the ordinary Bayes posterior density to that power $b$, as opposed to using $b$ as the learning rate in a generalized posterior.
\end{enumerate} 

The key point is that none of these can reliably be tuned to achieve calibration. First, the ordinary Bayes predictive in Item~1 has no free tuning parameters, so it will only be calibrated when the model is correctly specified. Second, adjusting the divergence measure with respect to which the predictive density referred to in Item~4 above is the Bayes estimator will provide no calibration guarantees. Third, the proposal in Item~2 has similar issues because the generalized posterior will, under certain conditions, concentrate around $\theta^\dagger$ when $n$ is large, which implies that the $(a,b,c)=(1,\eta,1)$ predictive density would be roughly $p_{\theta^\dagger}$. Since this is afflicted by model misspecification bias, i.e., $P_{\theta^\dagger} \neq P^\star$, and has no parameters left to be tuned, it cannot be successfully calibrated.  

The proposal in Item~3 above would avoid the criticism that model misspecification bias remains when $n$ is large. However, there is an even simpler proposal that can accomplish the same thing. Indeed, consider $a=\eta$ and $b=c=1$ in the above framework. That is, define the $\eta$-generalized predictive distribution as 
\begin{equation}
\label{eq:gpred}
f_n^{(\eta)}(y) \propto \int p_\theta^\eta(y) \, \Pi_n(d\theta), 
\end{equation}
where $\Pi_n=\Pi_n^{(1)}$ is the ordinary Bayes posterior. In the case where $\Pi_n$ concentrates around $\theta^\dagger$ for large $n$, it is clear from the expression in \eqref{eq:gpred} that we end up with $f_n^{(\eta)} \propto p_{\theta^\dagger}^\eta$, approximately. Since the dependence on a tunable $\eta$ remains, even with $n \to \infty$, we still have the flexibility to calibrate the predictive distribution. The key question is how can $\eta$ be tuned in order to achieve at least approximate calibration, and we address this question in Section~\ref{SS:gprc} below.  

To see that \eqref{eq:gpred} corresponds to a well-defined density, note that typically the learning rate would be between 0 and 1. By Jensen's inequality, if $\eta < 1$, then
\[ \int p_\theta^\eta(y) \, \Pi_n(d\theta) \leq \Bigl\{ \int p_\theta(y) \, \Pi_n(d\theta) \Bigr\}^\eta < \infty, \quad \text{for (almost) all $y$}, \]
and the integral in the upper bound above is simply the ordinary Bayes predictive density, say, $p_n$. If $p_n$'s tails are not too heavy, then $y \mapsto p_n^\eta(y)$ would be integrable, hence the right-hand side of \eqref{eq:gpred} defines a proper predictive distribution.  Note that the ``not too heavy'' condition concerns only the model---not the true $P^\star$---so it is entirely within the data analyst's control and can be readily checked in specific examples.  

Before we move on to the calibration algorithm, we should comment on the choice to work with the predictive distribution in \eqref{eq:gpred} as opposed to the one mentioned in Item~3 above that works with the $\eta$-generalized posterior $\Pi_n^{(\eta)}$. Recall that the primary role played by $\eta$ was to control the spread of the generalized posterior, with small $\eta$ leading to wider spread.  This is what motivated \citet{syring2019calibrating} to tune $\eta$ so that the nominal coverage could be achieved---the covariance mismatch, due to misspecification, could be (conservatively) overcome by stretching the posterior's contours sufficiently far.  In the prediction setting, however, we integrate over $\theta$ with respect to $\Pi_n$, so the shape of its contours is less important. That is, in prediction, the covariance mismatch has a $o(1)$ effect in the sense that $\Pi_n$ concentrates at $\theta^\dagger$ as $n \to \infty$, regardless of any (reasonable) adjustments that may have been made to the posterior contours. On the other hand, misspecification in the model density $p_{\theta^\dagger}$ remains, even asymptotically, so its effect is $O(1)$. Given that the generalized predictive in \eqref{eq:gpred}, with the ordinary posterior that ignores the $o(1)$ effect of covariance mismatch, is much simpler (see Section~\ref{SS:gprc}), we opt for this instead of the more complicated method described in Item~3 above.

\subsection{The GPrC algorithm}
\label{SS:gprc}

From the generalized predictive distribution $f_n^{(\eta)}$ in \eqref{eq:gpred}, with a particular $\eta$ value, we obtain an upper prediction limit as the solution $q=Q_\alpha(\eta; Y^n)$ of the equation
\[ \int_{-\infty}^q f_n^{(\eta)}(y) \, dy = 1-\alpha. \]
That is, $Q_\alpha(\eta;Y^n)$ is the upper-$\alpha$ quantile of the $\eta$-generalized predictive distribution. The goal is to select a value of the learning rate $\eta$ so that \eqref{eq:calibration} holds for $Q_\alpha(\eta; Y^n)$. Of course, all of what follows can be modified in the obvious way if a prediction lower limit or a prediction interval is desired instead of a prediction upper limit. 

Towards this, define the coverage probability of the prediction upper limit:
\[ c_\alpha(\eta) = P^\star\{Q_\alpha(\eta; Y^n) \geq Y_{n+1}\}. \]
Note that this is a probability with respect to the joint distribution of $(Y^n,Y_{n+1})$ under $P^\star$. The fact our notation $Q_\alpha$ includes ``$\alpha$'' does not guarantee that $Q_\alpha(\eta; Y^n)$ is a valid prediction upper limit; calibration in the sense of \eqref{eq:calibration} is what needs to be shown. That is, we aim to find $\eta$ to solve the equation 
\begin{equation}
\label{eq:learninrateeq}
c_\alpha(\eta) = 1-\alpha. 
\end{equation}
If $P^\star$ were known, then the coverage probability function $\eta \mapsto c_\alpha(\eta)$ could at least be evaluated numerically, to any desired accuracy, using Monte Carlo, and then the equation \eqref{eq:learninrateeq} could be solved using stochastic approximation (see below). In practice, however, $P^\star$ is unknown so a different strategy is required. Here we will make use of the bootstrap (in one form of another) to approximate the $P^\star$ probabilities using the observed data.  

To see more clearly where we are going, it may help to re-express the coverage probability using the familiar iterated expectation formula:
\begin{equation}
\label{eq:decomp}
c_\alpha(\eta) = E^\star \bigl[ P^\star\{Y_{n+1} \leq Q_\alpha(\eta; Y^n) \mid Y^n\} \bigr], 
\end{equation}
where $E^\star$ is expectation with respect to $P^\star$. This reveals that there are effectively two expectations that need to be approximated: one is over $Y_{n+1}$ with $Y^n$ fixed, and the other is over $Y^n$. Our approximations of these two expectations will be easiest to describe in the case of iid data; we will extend the idea to non-iid cases in Section~\ref{S:more} below. 

Let $\{Y_b^n\}_{b=1}^B$ be the $B$ bootstrap samples, each of size $n$, generated from $Y^n$. That is, each $Y_b^n$ is a random sample of size $n$, with replacement, from the observed data $Y^n$. Next, for each $b=1,\ldots,B$, let $Q_\alpha(\eta; Y_b^n)$ denote the quantile of the $\eta$-generalized predictive distribution in \eqref{eq:gpred} based on data $Y_b^n$; more details about this predictive quantile computation below. In the iid case, we have 
\[ P^\star(Y_{n+1} \leq x \mid Y^n) = P^\star(Y_{n+1} \leq x), \]
and the right-hand side can be readily estimated using the empirical distribution function from $Y^n$. This immediately leads to an empirical version of the expression in \eqref{eq:decomp}, 
\[ \hat c_\alpha(\eta) = \frac{1}{B} \sum_{b=1}^B \Bigl[ \frac{1}{n} \sum_{i=1}^n 1\{ Y_i \leq Q_\alpha(\eta; Y_b^n)\} \Bigr], \]
where $1\{A\}$ denotes the indicator function of the event $A$. For the iid case considered here, the inner average over $i=1,\ldots,n$ approximates the conditional probability in \eqref{eq:decomp}, given $Y^n$, and the outer average over $b=1,\ldots,B$ approximates the outer expectation with respect to the distribution of $Y^n$. 

The idea behind the GPrC algorithm is to solve $\hat c_\alpha(\eta) = 1-\alpha$ for $\eta$ instead of \eqref{eq:learninrateeq}, which leads to a data-driven choice, $\hat\eta$, of the learning rate $\eta$. To properly solve the equation $\hat c_\alpha(\eta)=1-\alpha$, we need a root-finding procedure that accommodates the Monte Carlo variability in $\hat c_\alpha$. As in \citet{syring2019calibrating}, we adopt the stochastic approximation method of  \citet{robbins1951stochastic}; see, also, \citet{kushner2003stochastic}. 
In particular, for a vanishing, deterministic sequence $\{\kappa_t: t \geq 1\}$, such that 
\[ \sum_{t=1}^\infty \kappa_t = \infty \quad \text{and} \quad \sum_{t=1}^\infty \kappa_t^2 < \infty, \]
and a starting value $\eta^{(1)}> 0$, define the sequence of candidate solutions 
\begin{equation}
\label{eq:eta.update}
\eta^{(t+1)} = \eta^{(t)} + \kappa_t \{ \hat c_\alpha(\eta^{(t)}) - (1-\alpha)\}, \quad t \geq 1. 
\end{equation}
If, instead of bootstrap samples, we could approximate the coverage probability function $c_\alpha$ using Monte Carlo samples from $P^\star$, then it could be checked using the standard convergence theory for stochastic approximation sequences (e.g., \citet{robbins1971convergence} that $\eta^{(t)}$ converges $P^\star$-almost surely to a solution of the equation in \eqref{eq:learninrateeq}. Given that bootstrap is a generally reliable computational tool for approximating sampling distributions, and that there is no reason to expect our present setup to be atypical (see Section~\ref{SS:more} below), we propose to update $\eta^{(t)}$ according to the rule \eqref{eq:eta.update} until it converges, and we denote the limit as $\hat\eta$. This makes up what we call the {\em generalized predictive calibration}, or {\em GPrC}, procedure; see Algorithm~\ref{alg:gprc}.

\begin{algorithm}[t]
\SetAlgoLined
\SetKwInOut{Input}{Input}
\SetKwInOut{Output}{Output}
\Input{Learning rate starting value,  $\eta^{(0)}=0.5$, \newline 
    the target quantile/coverage probability, $1-\alpha$,\newline
    $B$ bootstrap samples of size $n$, $\{Y_b^n: b=1,\dots,B\}$,\newline
    convergence tolerance, $\epsilon=\alpha\times0.01$ \newline 
    iteration index, $t=0$.}
\Output{A learning rate $\eta$ that calibrates the generalized predictive distribution in terms of a desired coverage probability.}
    
 \While{$|\hat{c}_\alpha(\eta^{(t)})-(1-\alpha)|>\epsilon$}{
    $\eta^{(t+1)} \gets \eta^{(t)}+\kappa_t\{\hat{c}_\alpha(\eta^{(t)})-(1-\alpha)\}$\;
   \For{$b=1,\dots,B$}{
    Estimate the upper-$\alpha$ quantile of each bootstrap sample, $Q_\alpha(\eta^{(t+1)}; Y^n_b)$\;
    }
   Evaluate $\hat{c}_\alpha(\eta^{(t+1)})$ using bootstrap samples\; 
   $t \gets t + 1$\;
 }
 \caption{Generalized Predictive Calibration (GPrC)}
\label{alg:gprc}
\end{algorithm}

In our implementation of the GPrC algorithm, we recommend a starting value $\eta^{(0)}=0.5$. The idea behind this choice is that we are anticipating some degree of model misspecification, in which case calibration would require $\eta < 1$, so we want to take a ``warm start'' in order to accelerate convergence. Other choices of starting values perform similarly, however. The convergence tolerance $\epsilon = 0.01 \times \alpha$ is intended to balance the quality of the coverage probability approximation versus the speed of convergence. We suggest using a cutoff that is an increasing function of $\alpha$ because calibrating at the extreme quantiles---small $\alpha$ values---is more challenging, hence a smaller tolerance is recommended in order to encourage more iterations. The convergence tolerance should also depend on the precision of the estimated coverage probabilities. In the present case of iid data, $\hat c_\alpha(\eta)$ is evaluated as an average of $nB$ indicators, so there is no point to make the tolerance $\epsilon$ less than $(nB)^{-1}$. But for the dependent data cases discussed later, there is less information available in the data, so the number of indicators being averaged to evaluate $\hat c_\alpha(\eta)$ is much smaller, hence less precision. In the spatial case, for example, the precision is bounded by $B^{-1}$, so we recommend a tolerance value of $\epsilon = \max\{0.01 \times \alpha, B^{-1}\}$. 

It remains to say a few words about the computation of $Q_\alpha(\eta; Y^n)$ for a given $\alpha$, $\eta$, and data set $Y^n$. The situations we have in mind (see the subsequent sections) are those where a simple model $\model$ is used with the GPrC algorithm there to correct for any misspecification bias. When the posited model is relatively simple, it may be possible to evaluate the posterior distribution and, hence, the ordinary Bayes predictive distribution in closed-form, e.g., if the prior is conjugate. In that case, evaluating the quantile $Q_\alpha(\eta; Y^n)$ can be solved using standard numerical methods. For more complicated models, Monte Carlo-based methods may be needed to evaluate the quantiles. Assuming one can obtain samples $\{\theta^{(m)}: m=1,\ldots,M\}$ from the posterior distribution $\Pi_n$, the $\eta$-generalized predictive density in \eqref{eq:gpred} can be approximated as 
\[ \hat f_n^{(\eta)}(y) \propto \frac1M \sum_{m=1}^M p_{\theta^{(m)}}^\eta(y), \]
and the normalizing constant and quantile can be found via quadrature. These Monte Carlo approximations would be required for each of the $B$ bootstrap samples, but not for each individual update of the learning rate $\eta$ in the GPrC algorithm.  This is where we find a computational advantage compared to the algorithm in \citet{syring2019calibrating}. In the latter reference, they are concerned with the posterior distribution, so the aforementioned covariance mismatch is crucial and cannot be ignored. As we explained above, the bias resulting covariance mismatch is a lower-order term in the prediction setting, and can be ignored. By not requiring the posterior to change with $\eta$, we can use the same $B$ sets of posterior samples in each of the $\eta$ updates in \eqref{eq:eta.update}. Therefore, updating $Q_\alpha(\cdot; Y_b^n)$ for a given bootstrap sample $b$ to reflect a change in the learning rate along the GPrC sequence is relatively inexpensive, so the computational cost is of the order $B$. Compare this to the original GPC algorithm in \citet{syring2019calibrating}, where an update of $\eta$ would have computational cost of the order $B \times M$, where $M$ is the desired number of Monte Carlo samples.  Therefore, our particular choice of the generalized posterior density in \eqref{eq:gpred} leads to a much faster and efficient algorithm than one which focuses on adjusting the learning rate in the posterior.  And in terms of computational time, in our examples here, at least for the iid setting in Section~\ref{S:result.simple} below, a run of GPrC takes only a matter of seconds to complete, even with $B=200$ or more.  


The GPrC algorithm can be extended to independent but not iid data cases, and even certain dependent data cases, with or without covariates. This only requires the use of suitable variations on the basic bootstrap approach described above designed to accommodate the assumed data structure. We make this extension for time series and spatial data applications in Sections~\ref{S:result.timeseries} and \ref{S:result.spatial}, respectively.

\section{Further details about GPrC}
\label{SS:more}

Here we highlight a few important features of our proposed procedure, in particular, what the tuned $\eta$-generalized predictive does and what values of $\eta$ lead to the key calibration property in \eqref{eq:calibration}. Simple, numerical examples will be given to illustrate these points. 

First, note that model misspecification bias cannot be corrected simply by adjusting a tuning parameter like $\eta$ in \eqref{eq:gpred}, at least not in general. So we have to be clear: we make {\em no claims} that there exists $\eta$ such that the $\eta$-generalized predictive distribution $f_n^{(\eta)}$ closely approximates $P^\star$ or, more generally, $P^\star(\cdot \mid Y^n)$, in any global sense. Since a complete correction of the model misspecification bias is generally out of reach, our goal instead is a more modest one ---to ensure that the predictive distribution achieves the calibration property \eqref{eq:calibration}, at least at a particular level $\alpha$, even when the model is misspecified; cf.~\citet{grunwald.safe}. This amounts to adjusting $\eta$ so that the tails of $f_n^{(\eta)}$ match those of $P^\star$ in a certain sense to be made clear below. 

\begin{figure}[t]
 \centerline{\includegraphics[scale=0.5]{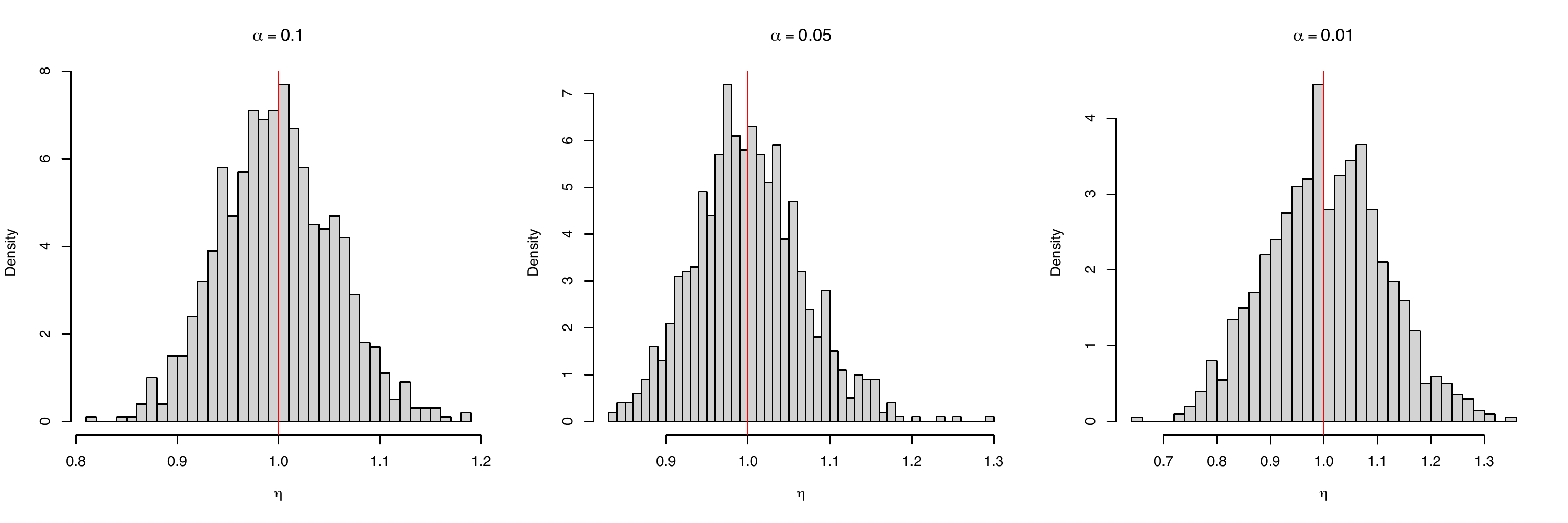}}
\caption{Histogram of the $\hat\eta$ values selected by the GPrC algorithm in the {\em correctly specified gamma example}, based on 1000 replications, each with sample size $n=400$.}
\label{fig:wellspecified_learningrate}
\end{figure}

The best place to start is with a correctly specified model. That is, let $P^\star = P_{\theta^\star}$ for some ``true'' parameter value $\theta^\star \in \Theta$, and let $p_{\theta^\star}$ denote the corresponding density. Using the asymptotic setting as a guide, recall that $f_n^{(\eta)}(y) \approx p_{\theta^\star}^\eta(y)$ when $n$ is large. Since the true distribution is calibrated at every level $\alpha$, we should take $\eta \approx 1$, which is what the GPrC algorithm does. For illustration, let $\gam(a,b)$ denote a gamma distribution with shape parameter $a > 0$ and rate parameter $b > 0$; the density function is 
\[ y \mapsto y^{a-1} e^{-b y}, \quad y > 0. \]
Consider a gamma model, $P_\theta= \gam(3,\theta)$, where the shape parameter is fixed at 3 but the rate parameter $\theta > 0$ is free to vary. The true value is $\theta^\star=2$ in our experiment.  With a conjugate gamma prior, $\theta \sim \gam(a,b)$, and iid data $Y^n=(Y_1,\ldots,Y_n)$, the posterior distribution is $\Pi_n = \gam(a_n, b_n)$, where $a_n = a + 3n$ and $b_n = b + \sum_{i=1}^n Y_i$. Finally, the $\eta$-generalized predictive distribution is the so-called {\em generalized beta prime distribution}  \citep[e.g.,][]{moghaddam2019generalized} with density 
\begin{equation}
\label{eq:beta.prime}
f_n^{(\eta)}(y) \propto \frac{(y/d)^{cp-1}}{\{1+(y/d)^c\}^{p+q}}, 
\end{equation}
where $(c,d,p,q)=(1, b_n/\eta, 2\eta+1, a_n+\eta-1)$.
For this illustration, we simulated 1000 data sets, each of size $n=400$, from the above gamma model, and ran the GPrC algorithm in each case to identify a learning rate $\hat\eta$ such that calibration at different $\alpha$ levels is achieved, at least approximately. Figure~\ref{fig:wellspecified_learningrate} shows the the distribution of the $\hat\eta$ values over the 1000 replications, at the three levels $\alpha \in \{0.01, 0.05, 0.10\}$.  Notice that the $\hat\eta$ values tend to concentration around $\eta=1$, as expected.   

Above we mentioned that the model misspecification bias {\em generally} cannot be corrected, i.e., there is no learning rate $\eta$ such that $f_n^{(\eta)}$ accurately approximates $P^\star$ in a global sense. There are, however, certain cases where the model misspecification bias is sufficiently mild that it can be completely corrected by tuning $\eta$. Suppose the true distribution is normal, $P^\star = \nm(\mu^\star, \sigma^{\star 2})$, and the model is $P_\theta = \nm(\theta, \sigma^2)$, where $\theta$ is unknown and to be inferred while $\sigma$ is fixed and generally different from $\sigma^\star$. This is relatively mild misspecification because the tails of the model basically match those of $P^\star$, so correcting for it may not be out of the question. With iid data $Y^n=(Y_1,\ldots,Y_n)$ and a conjugate prior $\theta \sim \nm(m, v)$, the posterior distribution is $\Pi_n = \nm(m_n, v_n)$, with 
\[ m_n = \frac{\sigma^2 m + v \sum_{i=1}^n Y_i}{\sigma^2 + n v} \quad \text{and} \quad v_n = \frac{\sigma^2 v}{\sigma^2 + n v}, \]
respectively. Then the $\eta$-generalized predictive distribution is also normal, i.e., $f_n^{(\eta)}$ is a $\nm(m_n, v_n + \eta^{-1} \sigma^2)$ density. When $n$ is large, $m_n \approx \mu^\star$ and $v_n \approx 0$, so to achieve calibration, we would need $\eta \approx (\sigma/\sigma^\star)^2$. Table~\ref{t:toy} compares the $\hat\eta$ values selected by the GPrC algorithm with $\eta^\dagger = (\sigma/\sigma^\star)^2$, for different values of $n$ and $\alpha$. Clearly, GPrC is tending to select $\hat\eta$ values near $\eta^\dagger$. Moreover, the empirical coverage probabilities of the $\hat\eta$-generalized predictive distribution quantiles shown in Table~\ref{t:toy} are all near the nominal level, suggesting that calibration in the sense of \eqref{eq:calibration} is achieved, at least approximately, across all $n$ and $\alpha$. Therefore, the GPrC algorithm successfully corrects for the model misspecification bias, which is relatively mild in this case.  

\begin{table}[t]
\begin{center}
\begin{tabular}{cccccc}
 \hline
$1-\alpha$ & $(\sigma/\sigma^\star)^2$   & $n=100$ & $n=200$& $n=400$ & $n=800$\\
 \hline
0.90 & 1 & 1.035 (0.885) & 1.016 (0.874) & 1.011 (0.890) & 0.998 (0.902) \\
& 0.8 & 0.816 (0.906) &   0.807 (0.901) &  0.802 (0.884) & 0.800 (0.898) \\
& 0.6 & 0.609 (0.895) & 0.607 (0.892) & 0.603 (0.885) & 0.599 (0.896) \\
& 0.4 & 0.441 (0.901) & 0.409 (0.881) & 0.404 (0.897) & 0.405 (0.903) \\
\hline
0.95 & 1 & 1.031 (0.948) & 1.022 (0.936) &  1.009 (0.943) & 1.005 (0.950) \\
& 0.8 & 0.827 (0.956) & 0.814 (0.946) & 0.812 (0.953) & 0.801 (0.944) \\
& 0.6 & 0.609 (0.951) & 0.615 (0.947) & 0.602 (0.944) & 0.600 (0.944) \\
& 0.4 & 0.425 (0.952) & 0.411 (0.951) & 0.403 (0.951) & 0.403 (0.958) \\
\hline
0.99 & 1 & 1.032 (0.987) & 1.021 (0.991) & 1.010 (0.993) & 1.007 (0.986) \\
& 0.8 & 0.829 (0.983) & 0.817 (0.986) & 0.809 (0.994) & 0.802 (0.988) \\
& 0.6 & 0.626 (0.978) & 0.614 (0.992) & 0.609 (0.991) & 0.598 (0.988) \\
& 0.4 & 0.437 (0.985) & 0.420 (0.990) & 0.408 (0.992) & 0.408(0.990) \\
\hline
\end{tabular}
\end{center}
\caption{Average $\hat\eta$ values from the GPrC algorithm and (in parentheses) the empirical coverage probability of the corresponding $\hat\eta$-generalized predictive distribution quantiles, at three different levels $\alpha \in \{0.01, 0.05, 0.10\}$, in the {\em mildly misspecified normal example}, based on 1000 replications.}
\label{t:toy}
\end{table}

\begin{figure}[t]
\centerline{
\includegraphics[scale=0.5]{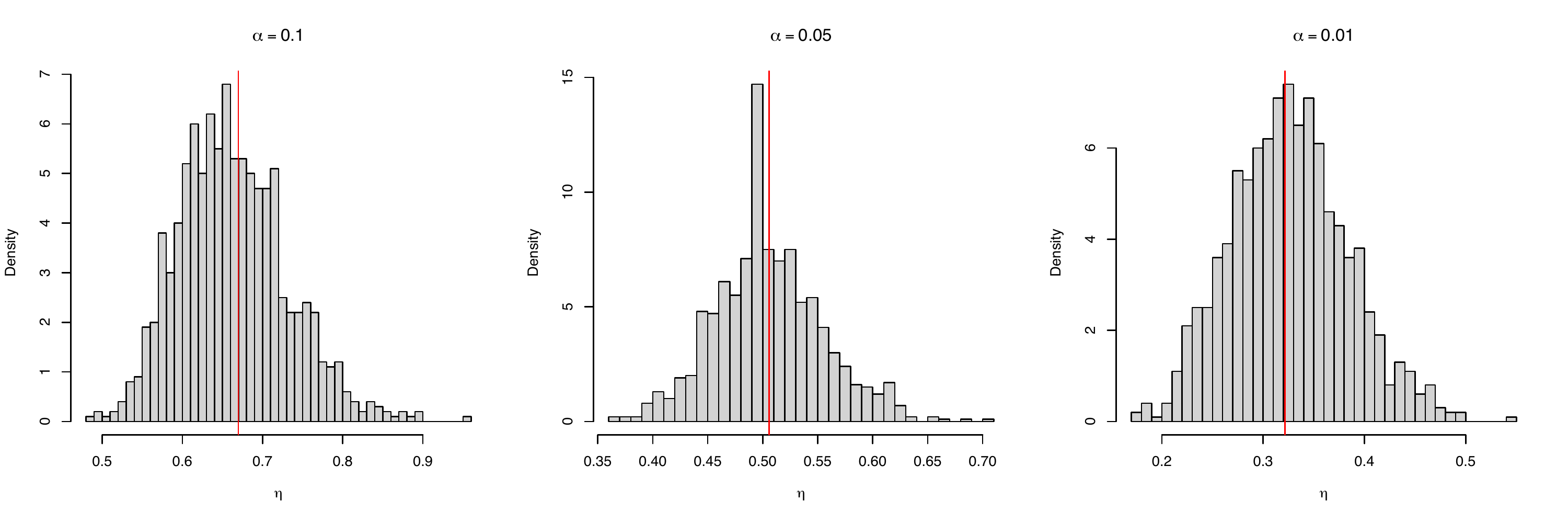}
}
\caption{Histogram of the $\hat\eta$ values selected by the GPrC algorithm across $1000$ replicates, each with $n=400$, in the {\em misspecified log-normal--gamma example}. The average learning rates that solve \eqref{eq:Q.sol} across the replications are shown by red vertical lines at 0.670, 0.506, and 0.322 in the $\alpha=0.10$, $\alpha=0.05$, and $\alpha=0.01$ cases, respectively.}
\label{fig:gamma.logn.1}
\end{figure}

Next, what do we expect the GPrC algorithm to do under more severe model misspecification? For example, suppose that we use the same gamma model described above but it happens that the true distribution is log-normal, say, $Y^n$ consists of iid observations coming from $\logn(\mu^\star, \sigma^{\star 2})$, with $\mu^\star = \sigma^\star = 1$. Based on how the GPrC algorithm is defined, we would expect that it would choose $\eta$ so that the upper-$\alpha$ quantile of the $\eta$-generalized predictive distribution given in \eqref{eq:beta.prime} agrees with the upper-$\alpha$ quantile of the true distribution, $\logn(\mu^\star, \sigma^{\star 2})$. That is, GPrC is aiming to solve the equation 
\begin{equation}
\label{eq:Q.sol}
Q_\alpha(\eta; Y^n) = Q_\alpha^\star, 
\end{equation}
where $Q_\alpha^\star=\exp\{\mu^\star+\sigma^\star \, \Phi(1-\alpha)\}$.  In practice, however, GPrC can only find an approximate solution, because an exact solution would require knowledge of $P^\star$ or, in this case, $(\mu^\star,\sigma^\star)$, which is information the algorithm does not have.  But this is a simulation study, where $P^\star$ is known, so it is possible to solve the equation \eqref{eq:Q.sol} exactly.  We did precisely this and the results are summarized in Figure~\ref{fig:gamma.logn.1}. That is, for each of 1000 replications, with sample size $n=400$, we evaluate both the $\hat\eta$ value from the GPrC algorithm and the solution of the equation \eqref{eq:Q.sol}, for $\alpha \in \{0.01, 0.05, 0.10\}$, and the plot shows histograms of the former compared to the average (red vertical line) of the latter. The key observation is that the GPrC estimates are centered around the average of those solutions to the equation \eqref{eq:Q.sol}, which confirms our claim that GPrC aims to match the quantile of the $\eta$-generalized predictive distribution to that of the true distribution.  

For a closer look at the GPrC algorithm's performance, we consider comparisons with three other methods. The first, denoted by {\em gamma}, is where we stick with the ordinary Bayes predictive distribution, with $\eta \equiv 1$, under the misspecified model. The second is the proposed method, denoted by {\em gamma + GPrC}, where the misspecified gamma model is assumed but the learning rate $\eta$ chosen according to GPrC. Third, is a (version of the) Bayesian nonparametric formulation based on a Dirichlet process mixture model, which we denote here by {\em DP Mixture}. The variation being employed here is the fast, recursive approximation, originally motivated by  \citet{newton1998nonparametric} and  \citet{newton2002nonparametric}---see, also, \citet{tokdar2009consistency} and  \citet{martin2019survey}---and developed fully for the prediction setting in \citet{hahn2018recursive}. Finally, the last method, denoted by {\em log-normal}, is not really a method, it is the oracle that uses the true distribution for prediction. 

We simulate data from the log-normal model and compare the coverage probability of the $100(1-\alpha)$\% prediction upper limits or, equivalently, the upper-$\alpha$ quantiles of the various predictive distributions. Note that, despite the non-trivial model misspecification bias, the gamma + GPrC is able to calibrate its prediction limits, while the simple gamma model cannot. In this case, the DP mixture method too is able to successfully calibrate its prediction limits; but see Section~\ref{S:result.simple}.  


\begin{table}[t]
\begin{center}
\begin{tabular}{cccc}
 \hline
Model   & $\alpha=0.10$ & $\alpha = 0.05$ & $\alpha=0.01$ \\
 \hline
Gamma & 0.872 & 0.876 & 0.939 \\
Gamma + GPrC & 0.908 & 0.943 & 0.989\\
DP Mixture & 0.911 & 0.947 & 0.992\\
Log-normal & 0.914 & 0.945 & 0.994\\
\hline 
\end{tabular}
\end{center}
\caption{Empirical coverage probabilities for the upper $100(1-\alpha)$\% prediction limits, for four methods in the {\em misspecified log-normal--gamma} example, based on 1000 replications.}
\label{t:one}
\end{table}


We conclude this section with one last example that is simple enough to do the relevant calculations in closed-form. Suppose that the model $P_\theta$ is $\nm(\mu,\sigma^2)$, where $\theta=(\mu,\sigma^2)$, but that the true distribution $P^\star$ is a Laplace distribution, denoted by $\lap(\mu^\star, \lambda^\star)$, where $\mu^\star$ is the mean and $\lambda^\star$ is the scale parameter; that is, $P^\star$ has density 
\[ p^\star(y) = (2\lambda^{\star})^{-1} e^{-|y-\mu^\star|/\lambda^\star}, \quad y \in \RR. \]
For a Bayesian analysis, we proceed by introducing a conjugate normal--inverse gamma prior for $(\mu, \sigma^2)$, where the conditional prior for $\mu$, given $\sigma^2$, is $\nm(m, k\sigma^2)$ and the marginal prior for $(1/\sigma^2)$ is $\gam(a,b)$. The prior hyperparameters, $(m,k,a,b)$, are taken to be fixed constants. Then it is not too difficult to show that the $\eta$-generalized predictive density \eqref{eq:gpred} is a location-scale transformation of a Student-t density; that is, 

\[ (Y_{n+1} \mid Y^n,\eta) \sim m_n + \Bigl\{ \Bigl( \frac{1}{\eta} + \frac{k}{nk+1} \Bigr) \frac{2b_n}{2a_n+\eta-1} \Bigr\}^{1/2} \, \stt_{2a_n+\eta-1}, \]
where 
\[ m_n = \tfrac{nk}{nk+1} \, \hat\mu_n + \tfrac{1}{nk + 1} \, m, \qquad a_n = a + \tfrac{n}{2}, \qquad 
b_n = b + \tfrac{n}{2} \hat\sigma_n^2 + \tfrac{n}{2(nk+1)} (\hat\mu_n - m)^2, \]
with $\hat\mu_n = n^{-1} \sum_{i=1}^n Y_i$ and $\hat\sigma_n^2 = n^{-1} \sum_{i=1}^n (Y_i - \hat\mu_n)^2$. Since the mean and variance of $\lap(\mu^\star, \lambda^\star)$ are $\mu^\star$ and $2\lambda^{\star 2}$, respectively, we find that $\hat\mu_n \to \mu^\star$ and $\hat\sigma_n^2 \to 2\lambda^{\star 2}$ in $P^\star$-probability. Since $a_n \to \infty$, it follows that, for large $n$, the $\eta$-generalized predictive density can be approximated by 
\[ f_n^{(\eta)}(y) \approx \nm(y \mid \mu^\star, 2\eta^{-1} \lambda^{\star 2}). \]
Incidentally, the minimizer of the Kullback--Leibler divergence of $P_\theta$ from $P^\star$ is $\theta^\dagger = (\mu^\star, 2\lambda^{\star 2})$, so the right-hand side of the above approximation agrees with that mentioned above based on the posterior concentration properties of $\Pi_n$ under model misspecficiation, i.e., $f_n^{(\eta)}(y) \propto p_{\theta^\dagger}^\eta(y)$, approximately. Since the upper-$\alpha$ quantile of $\lap(\mu^\star, \lambda^\star)$ is 
\[ Q_\alpha^\star = \mu^\star - \lambda^\star \log(2\alpha), \]
using the above approximation, we find that $\eta$ should be chosen such that 
\[ \mu^\star + (2\eta^{-1} \lambda^{\star 2})^{1/2} \, \Phi^{-1}(1-\alpha) = Q_\alpha^\star, \]
or, equivalently, we want to use 
\[ \eta_\alpha = 2 \{\Phi^{-1}(1-\alpha) / \log(2\alpha)\}^{2}. \]
Note, again, that this ``ideal'' choice of $\eta$ depends on $\alpha$; however, in this case, since both the model and the true distribution are location-scale families, $\eta_\alpha$ does not depend on features of $P^\star$, so this value is accessible in real applications.  

\begin{figure}[t]
 \centerline{\includegraphics[scale=0.6]{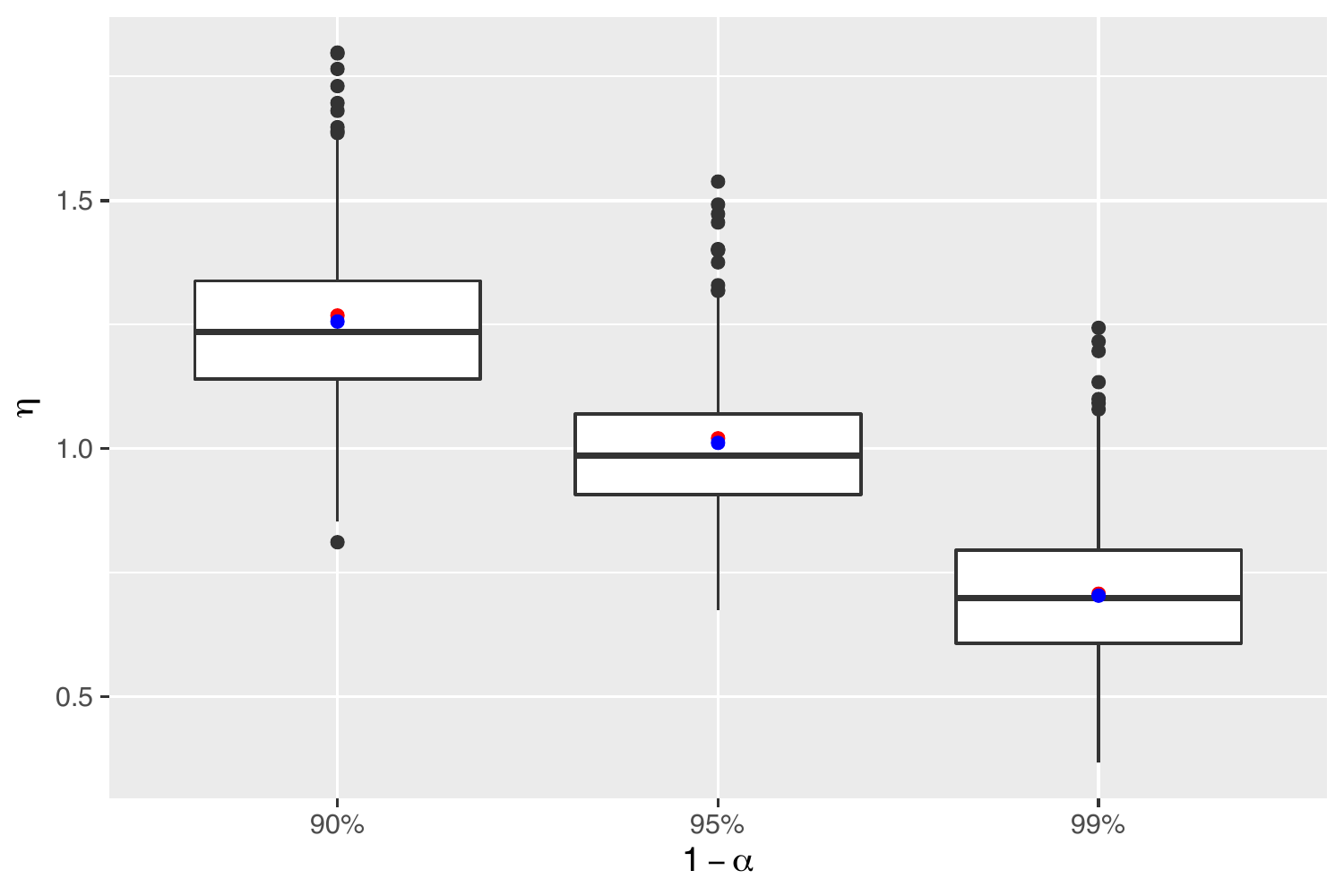}}
\caption{Boxplot of the learning rate selected by the GPrC algorithm in the {\em misspecified Laplace--normal} example  $(\mu^\star=0,\lambda^\star=1)$, based on $1000$ replications, each with $n=400$.  Red dots denote $\eta_\alpha$ and blue dots denote the average $\eta$ to match the predictive distribution quantiles to $Q_\alpha^\star$.}
\label{fig:Laplace_example_learningrate}
\end{figure}

To confirm that, indeed, the GPrC algorithm selects $\eta$ close to the $\eta_\alpha$ described above, we do a brief simulation study. The boxplots in Figure~\ref{fig:Laplace_example_learningrate} summarize the $\hat\eta$ values chosen by the GPrC algorithm, the red dots correspond to the $\eta_\alpha$ value defined above, and the blue dots correspond to the average $\eta$ value obtained matching the actual, data-dependent $\eta$-generalized predictive distribution---not the asymptotic approximation---to the true quantile $Q_\alpha^\star$. As expected, the red and blue dots are indistinguishable, and the GPrC algorithm's learning rate choices tightly concentrate around these ``ideal'' $\eta$ values, confirming its effectiveness. Interestingly, note that at least for moderate $\alpha$ levels, the ideal $\eta_\alpha$ and those that tend to be chosen by GPrC are larger than 1. The reason is that the Student-t predictive density---but not the normal limit---is {\em wider} than the true Laplace density, so GPrC becomes more aggressive, adaptively shrinking the prediction intervals while maintaining calibration. 

\section{Illustrations: skewed, heavy-tailed cases}
\label{S:result.simple}

\subsection{Setup and take-away messages}
\label{SS:sims}

Our original motivation behind the GPrC algorithm was that in, e.g., actuarial science of finance applications, often the goal is to predict observations when the underlying distribution is skewed and heavy-tailed. It can be difficult to specify good, finite-dimensional parametric models to handle such data; moreover, nonparametric methods are more complicated and the most commonly used versions---Dirichlet process mixtures of normal kernels with a thin-tailed base measure for the prior---are suited only for cases where the tails are not too heavy. Therefore, it would be beneficial if one could take a simple, pragmatic model, which is not assumed to be correctly specified, and let the data determine what kind of adjustments (if any) are needed. So, for our first set of illustrations, we consider heavy-tailed data, like what often manifests in financial applications, and use the GPrC algorithm to adjust so that the predictive distribution is calibrated. In particular, we consider two different $P^\star$s, namely, the Pareto and generalized extreme value distributions. Both are supported on $[0,\infty)$, hence are skewed, and have a shape parameter that controls the heaviness of the tail; the specific distributional forms are given below. In both cases, we consider a simple log-normal model, $P_\theta = \logn(\mu, \sigma^2)$, with a conjugate prior for $\theta=(\mu, \sigma^2)$. Since log-normal has relatively thin tails, it can be severely misspecified depending on the Pareto or generalized extreme value distribution's shape parameter, so makes for a good test of the GPrC methodology.  

We compare the GPrC results with the misspecified Bayesian predictive distributions and the aforementioned variation on the Bayesian nonparametric Dirichlet process mixture model. The comparison will be based on two metrics, namely, empirical coverage probability of $100(1-\alpha)$\% upper prediction limits and a one-sided version of the empirical interval score suggested by \citet{gneiting2007strictly}. To make this precise, we need a bit more notation. Let $R$ denote the number of replications, which we take to be $R=1000$ in our experiments. For each $r=1,\ldots,R$, we simulate data $(Y_r^n,Y_{r,n+1})$ from $P^\star$, and obtain the prediction upper limit $\widehat Q_\alpha(Y_r^n)$ based on each method. Then the empirical coverage probability is defined as 
\[ \frac1R \sum_{r=1}^R 1\{\widehat Q_\alpha(Y_r^n) \geq Y_{r,n+1}\}. \]
Of course, the empirical coverage probability should be close to the nominal level $1-\alpha$.  Similarly, we consider a relative interval score $S/S^\star$, where $S$ is given by 
\[ S = \frac1R \sum_{r=1}^R \bigl[ \widehat Q_\alpha(Y_r^n) + \alpha^{-1} \{ Y_{r,n+1} - \widehat Q_\alpha(Y_r^n)\} \, 1\{Y_{r,n+1} > \widehat Q_\alpha(Y_r^n)\} \bigr], \]
and $S^\star$ is the same but with the true quantile $Q_\alpha^\star$ of $P^\star$:
\[ S^\star = \frac1R \sum_{r=1}^R \bigl[ Q_\alpha^\star + \alpha^{-1} ( Y_{r,n+1} - Q_\alpha^\star) \, 1\{Y_{r,n+1} > Q_\alpha^\star\} \bigr]. \]
Small interval scores are better and, since the true quantile $Q_\alpha^\star$ is ``best'' in the sense of having the smallest interval score, we expect $S/S^\star \geq 1$ and closer to 1 is better. In our experiments we vary $n \in \{100, 200, 400\}$ and $\alpha \in \{0.01, 0.05, 0.10\}$.  

The take-away messages here are two-fold.  First, as expected, the greater the disparity between the posited model and true distribution, and further out in the tails of the distribution one aims to predict, the performance of a naive method that does not adjust for possible model misspecification gets worse.  In particular, the coverage probability of an unadjusted model-based prediction upper bounds can be well below the nominal level.  Second, our DP mixture does well overall in terms of coverage, but tends to be less efficient in terms of interval score compared to the GPrC algorithm.  Moreover, the GPrC solution can be readily extended to cases beyond the simple iid prediction problems considered here in this section, while the DP mixture formulation is far less straightforward.


\subsection{Pareto data}

\begin{figure}[p]
\begin{center}
\includegraphics[scale=0.52]{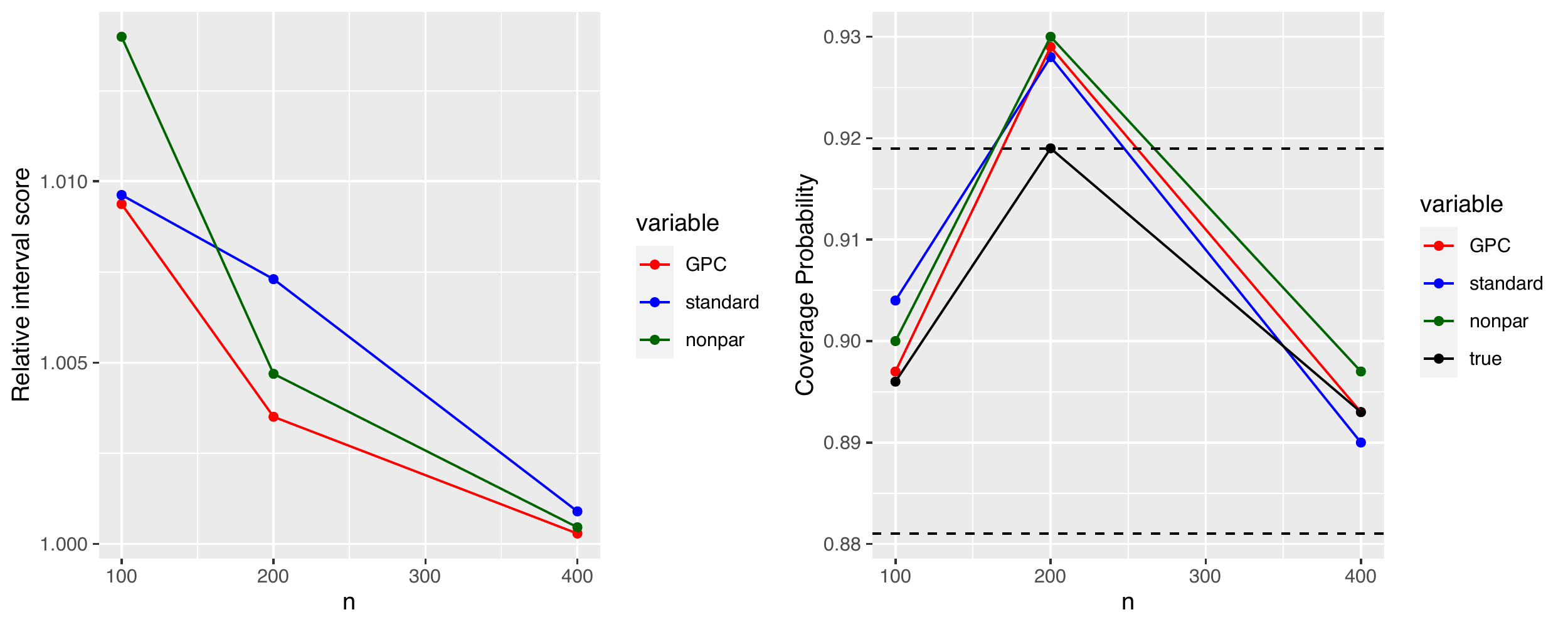}
\includegraphics[scale=0.52]{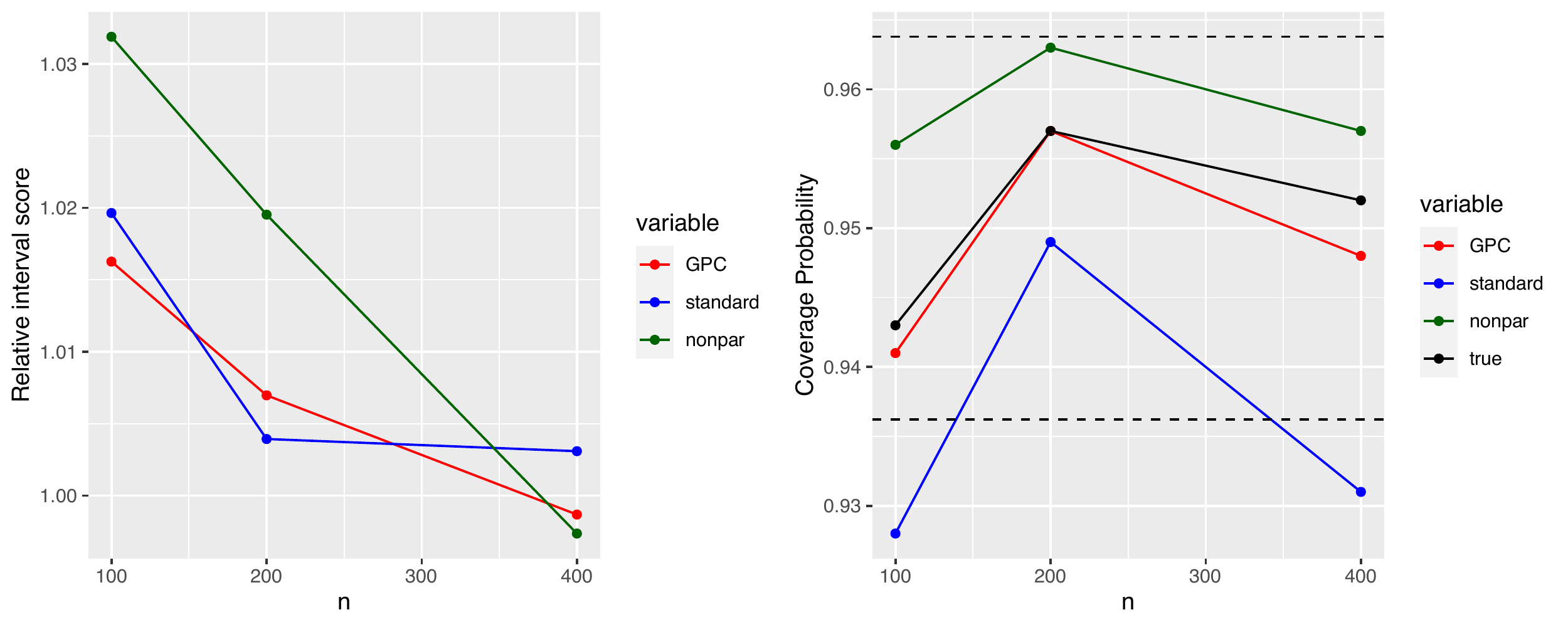}
\includegraphics[scale=0.52]{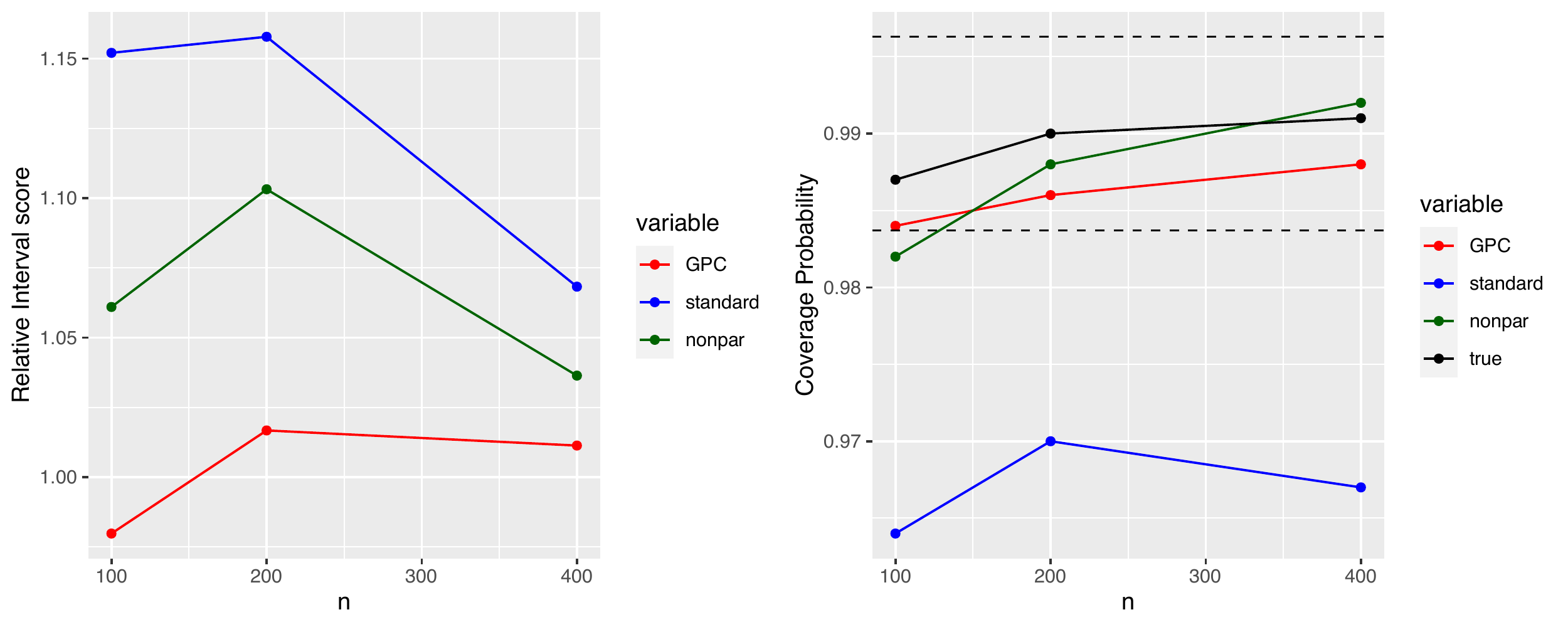}
\end{center}
\caption{Summary of prediction limit performance in the {\em Pareto--log-normal} example, with $a=2$. Left column shows the relative interval score; right column shows the coverage probability. Top to bottom, the rows correspond to $\alpha=0.10$, $\alpha=0.05$, and $\alpha=0.01$. Dashed horizontal lines in the right column correspond to two Monte Carlo standard errors around the target coverage probability $1-\alpha$.}
\label{fig:pareto}
\end{figure}

As a first illustration involving skewed, heavy-tailed data, suppose the true distribution $P^\star$ is Pareto, with distribution function given by 
\[ P^\star(Y \leq y) = 1 - (1 + y)^{-a}, \quad y \geq 0, \]
where $a > 0$ is {\em shape parameter} that controls the heaviness of the tails. In particular, smaller $a$ means the distribution function approaches 1 more slowly as $y \to \infty$, hence a heavier tail. So the ``degree of misspecification''---of the Pareto with shape parameter $a$ compared to log-normal---is increasing as $a$ decreases. In our experiments, we considered three such degrees of misspecification, namely, $a=4$, $a=3$, and $a=2$. In all cases, the standard Bayesian solution that ignores the potential model misspecification performs poorly, especially so in the case with the highest degree of misspecification. For the two moderate cases, $a \in \{3,4\}$, the log-normal + GPrC and the DP mixture methods perform similarly in terms of relative interval score and coverage probability, so we omit the detailed comparisons. We focus here on the case $a=2$ with highest degree of misspecification, which is arguably the most interesting.  Figure~\ref{fig:pareto} shows the relative interval scores and empirical coverage probabilities of the various $100(1-\alpha)$\% upper prediction limits, as functions of $n$, for various $\alpha$ levels. The take-away message is that both log-normal + GPrC and DP mixture are able to achieve the nominal coverage probability across $n$ and $\alpha$, but that the former does so with a slightly better relative interval score, suggesting a benefit in terms of overall interval efficiency. Clearly, the standard Bayes solution that ignores the misspecification is not competitive in this very heavy-tailed situation with a high degree of misspecification.  

\subsection{Generalized extreme value data}

For our second illustration involving skewed, heavy-tailed data, suppose $P^\star$ is a generalized extreme value distribution, with distribution function 
\[ P^\star(Y \leq y) = \exp\bigl[ -\{1 + \xi(y - 2)\}^{-1/\xi} \bigr], \quad y \geq 2 - \xi^{-1}, \]
where $\xi > 0$ is the shape parameter.  Like in the Pareto example above, the shape parameter controls the heaviness of the generalized extreme value distribution's tails.  However, in this case, heaviness of the tails, or the ``degree of misspecification,'' is increasing in $\xi$.  For our experiments we consider $\xi=0.2$, $\xi=0.5$, and $\xi=0.7$.  For the moderate degrees of misspecification, namely, $\xi=0.2$ and $\xi=0.5$, both log-normal + GPrC and DP mixture perform comparably, so we omit the detailed results and focus our attention on the most interesting case, $\xi=0.7$, corresponding to a very heavy-tailed $P^\star$, which makes the thin-tailed log-normal model significantly misspecified. 

\begin{figure}[p]
\begin{center}
\includegraphics[scale=0.52]{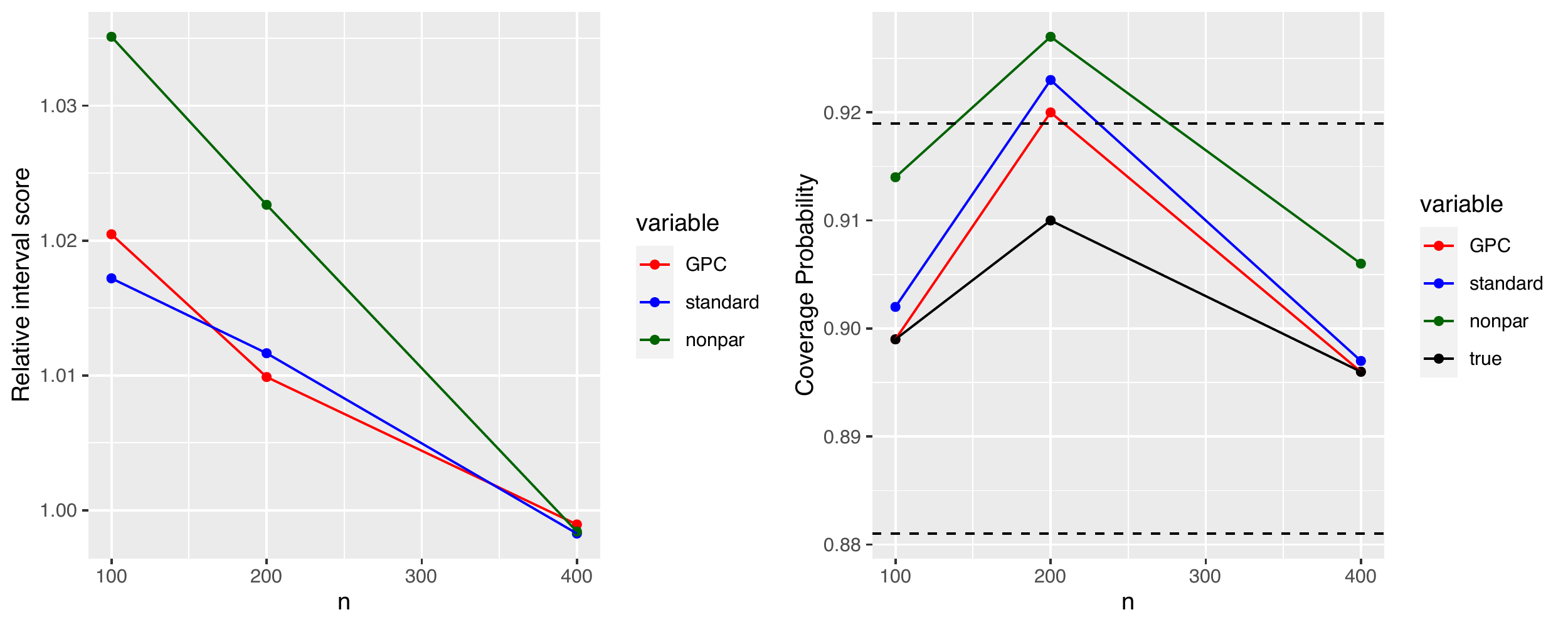}
\includegraphics[scale=0.52]{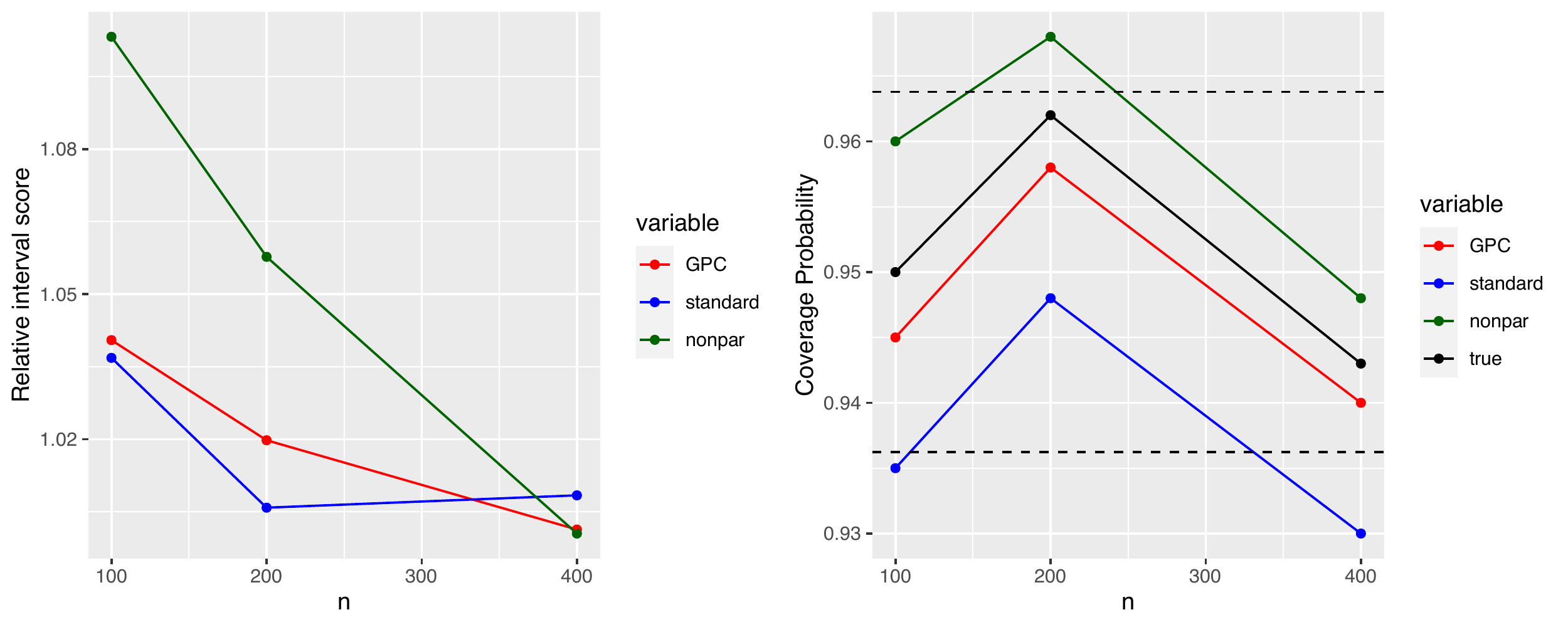}
\includegraphics[scale=0.52]{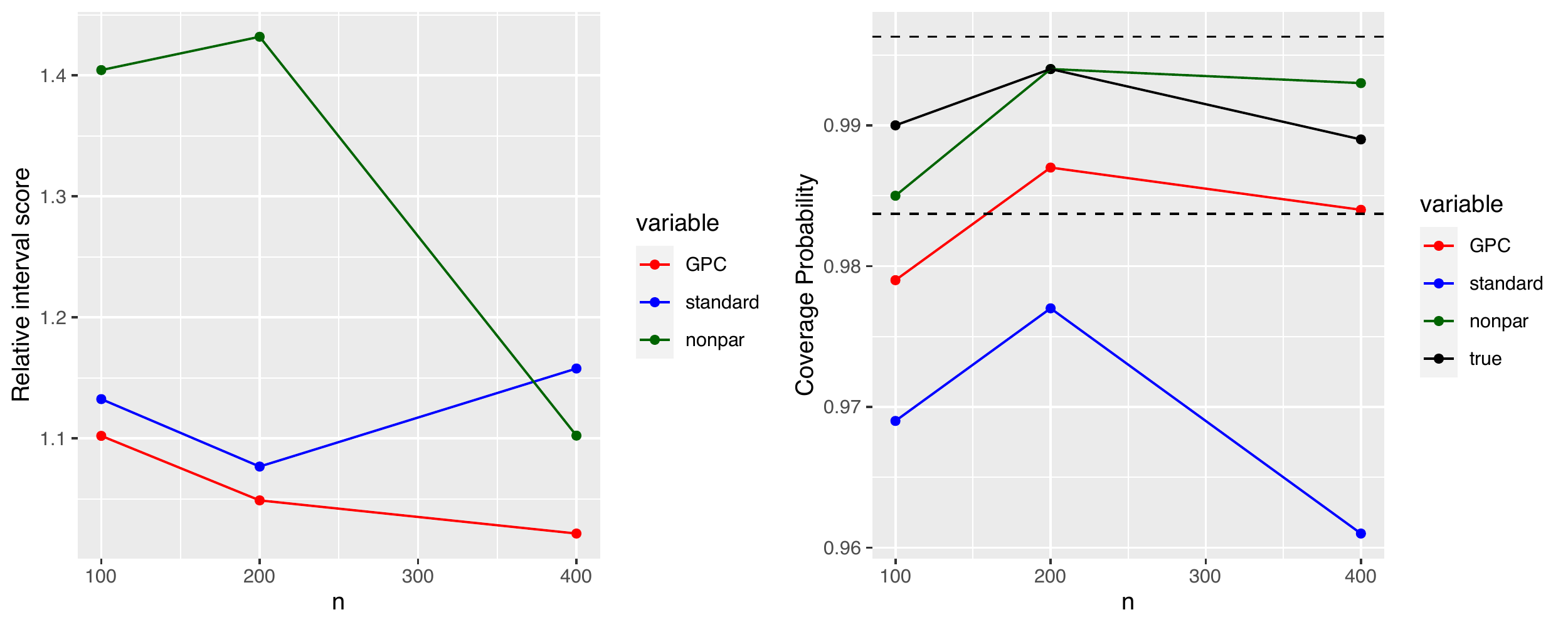}
\end{center}
\caption{Summary of prediction limit performance in the {\em generalized extreme value--log-normal} example, with $\xi=0.7$.  Left column shows the relative interval score; right column shows the coverage probability.  Top to bottom, the rows correspond to $\alpha=0.10$, $\alpha=0.05$, and $\alpha=0.01$. Dashed horizontal lines in the right column correspond to two Monte Carlo standard errors around the target coverage probability $1-\alpha$.}
\label{fig:gev}
\end{figure}

Like above, Figure~\ref{fig:gev} shows the relative interval scores and empirical coverage probabilities of the various $100(1-\alpha)$\% upper prediction limits, as functions of $n$, for various $\alpha$ levels.  In this case, at the less extreme $\alpha$ levels, namely, $\alpha=0.10$ and $\alpha=0.05$, both the Bayes solution that ignores misspecification and the log-normal + GPrC that adjusts for it perform well in terms of interval score and coverage.  The DP mixture method appears to be producing too wide of prediction intervals, as indicated by the large interval score. For the extreme quantile, $\alpha=0.01$, the separation between the methods becomes more clear and the benefits of GPrC's adjustments to specifically achieve coverage emerge, as we see by its ability to cover within an acceptable range of the nominal 99\% rate and have the smallest interval score.  

\section{Calibration in regression problems}
\label{S:regression}

Here we extend the GPrC methodology to cases where the $Y_i$'s are accompanied by predictor variables $X_i \in \RR^d$, for some $d \geq 1$.  Here we focus on the setting in which the pairs $D_i = (X_i, Y_i) \in \RR^{d+1}$ are iid; the setting in which the predictors are non-random can be handled similarly, and is discussed briefly at the end of this section. 

Let $P^\star$ denote the true  distribution of $D=(X,Y)$, and suppose that $D_i = (X_i,Y_i)$ are iid copies of $D$, for $i=1,\ldots,n$.  The data analyst would typically opt to model only the conditional distribution of $Y$, given $X=x$, by a distribution $P_{\theta;x}$, with a density $p_\theta(\cdot \mid x)$, which depends on a model parameter $\theta \in \Theta$.  The most common example of this, which we will adopt here, is the textbook linear regression model with 
\[ P_{\theta; x} = \nm\bigl( g_\beta(x), \sigma^2 \bigr), \quad \theta = (\beta,\sigma^2) \in \RR^{q+1}, \]
where $g_\beta: \RR^d \to \RR$ belongs to a given parametric class of functions indexed by $\beta \in \RR^q$; the common linear model corresponds to $g_\beta(x) = x^\top \beta$.  This choice to model only the conditional distribution is equivalent to assuming a joint model for $(X,Y)$ but assuming the marginal distribution for $X$ is known and does not depend on $\theta$.  In any case, a Bayesian approach can be carried out, which leads to a posterior distribution, $\Pi_n$, of $\theta$ that depends on data $D^n$.  The presence of the predictor variables makes the predictive distribution in this case is slightly different from before.  Indeed, the $\eta$-generalized predictive distribution here is 
\[ f_n^{(\eta)}(y \mid x) \propto \int p_\theta^\eta(y \mid x) \, \Pi_n(d\theta). \]
That is, our prediction of $Y_{n+1}$ depends on a value of the associated $X_{n+1}$, which is available to the data analyst at the time prediction is to be carried out.  

From this $\eta$-generalized predictive distribution comes an upper-$\alpha$ quantile, which we will denote by $Q_\alpha(\eta; x, D^n)$, which makes its dependence on the value $x$ of the predictor variable explicit.  Define the coverage probability as 
\[ c_\alpha(\eta) = P^\star\{ Q_\alpha(\eta; X_{n+1}, D^n) \geq Y_{n+1}\}, \]
and note that this probability is taken with respect to the true joint distribution of $D^{n+1} = \{(X_i,Y_i): i=1,\ldots,n+1\}$ under $P^\star$. The GPrC algorithm can be applied here in almost exactly the same way as in the basic iid setup in Section~\ref{SS:gprc}.  That is, take $B$ many bootstrap samples $D_b^n$ from $D^n$, which amounts to $B$ distinct samples of pairs $(X_i,Y_i)$ with replacement from the original sample---this is the so-called {\em paired bootstrap} \citep[e.g.,][]{efron1992bootstrap, freedman1981bootstrapping, flachaire2005bootstrapping, rabbi2020model}.  Then approximate the above coverage probability by 
\[ \hat c_\alpha(\eta) = \frac1B \sum_{b=1}^B \Bigl[ \frac1n \sum_{i=1}^n 1\{Q_\alpha(\eta; X_i, D_b^n) \geq Y_i\} \Bigr], \]
where each term in the inner sum is based on applying the $\eta$-generalized predictive distribution to predict $Y_i$ at the given $X_i$. Note that this formula does not actually use the value $X_{n+1}$.  That value would, however, be used for forming the actual predictive distribution, $f_n^{(\hat\eta)}(y \mid X_{n+1})$ and associated quantile $Q_\alpha(\hat\eta; X_{n+1}, D^n)$ that would be used to predict $Y_{n+1}$ in the real application.  We use $Q_\alpha(\hat\eta; D^n)$ in our simulations below to check the coverage probability of the proposed GPrC-based method. 

Here we consider linear regression model misspecification in terms of the distribution of the error terms. In particular, we consider two skewed error distributions, namely 
$\Chisq(2)$ and generalized extreme value distribution with shape $\xi=0.5$, both centered to have mean zero. Note that the latter is both skewed and heavy-tailed.  The rows of the $X$  matrix is sampled from a mean-zero multivariate normal distribution with a unit variance and first-order autoregressive structure, i.e., $\E(x_{ij} x_{ik}) = \rho^{|j-k|}$, with correlation $\rho=0.5$. We also use $\beta=(2,2,2,2,2)\top$ to sample $Y_i$, and compare the GPrC results with the plug-in predictive interval using maximum likelihood estimator with a standard normal error distribution.  Empirical coverage probabilities are presented in Tables~\ref{t:ChisqLRerror}--\ref{t:GEVLRerror}. Both GPrC and plug-in methods perform well in terms of coverage at $\alpha=0.10$, so these details are not shown. At the more extreme $\alpha$ levels, especially $\alpha=0.01$, the plug-in method's predictive intervals are too narrow to cover within an acceptable range of the 99\% target, even with a relatively large sample size.  


\begin{table}[t]
\begin{center}
\begin{tabular}{ccccc}
 \hline
$1-\alpha$ & Method  & $n=100$ & $n=200$& $n=400$ \\
 \hline
0.95 & GPrC & 0.941 & 0.957 &  0.963\\
& Plug-in & 0.919 & 0.939 & 0.930  \\
\hline
0.99 & GPrC & 0.991 & 0.989 & 0.988 \\
& Plug-in & 0.967 & 0.960 & 0.965\\
\hline
\end{tabular}
\end{center}
\caption{Empirical coverage probability of the corresponding $\hat\eta$-generalized predictive distribution quantiles, at two confidence levels $\alpha \in \{0.05, 0.01\}$, in the {\em centered chi-square error distribution example}, based on 1000 replications.}
\label{t:ChisqLRerror}
\end{table}

\begin{table}[t]
\begin{center}
\begin{tabular}{ccccc}
 \hline
$1-\alpha$ & Method  & $n=100$ & $n=200$& $n=400$ \\
 \hline
0.99 & GPrC & 0.988 & 0.987 & 0.990 \\
& Plug-in & 0.975 & 0.976 & 0.978\\
\hline
\end{tabular}
\end{center}
\caption{Empirical coverage probability of the corresponding $\hat\eta$-generalized predictive distribution quantiles, at confidence levels $\alpha =0.01$, in the {\em centered generalized extreme value error distribution example}, based on 1000 replications.}
\label{t:GEVLRerror}
\end{table}

Regression problems involving fixed/non-random covariates require a slightly different formulation.  Suppose that $Y^n = (Y_1,\ldots,Y_n)$ consists of independent observations, where $Y_i$ has an $i$-specific marginal distribution $P_{\theta; x_i}$, depending on a fixed covariate $x_i$.  The most common example of this is in a designed study where the $Y_i$ measurements are taken under pre-determined experimental settings.  Then the goal is to predict $Y_{n+1}$ under another pre-determined setting $x_{n+1}$.  The same construction of an $\eta$-generalized predictive distribution described above can be applied here.  The only difference in the GPrC formulation comes in the bootstrap approximation of the coverage probability: in this case, the paired bootstrap is replaced by the {\em residual boostrap}  \citep[e.g.,][]{efron1982jackknife}.

\section{Calibration in dependent data problems}
\label{S:more}

\subsection{Time series}
\label{S:result.timeseries}

Here we consider the problem of calibrating predictive distributions when the data are dependent.  We start here with the simplest case of time series---or temporally dependent---data; the next section considers spatially dependent data.  

Suppose we have data $Y^n = (Y_1,\ldots,Y_n)$ and the goal is to predict $Y_{n+1}$.  Naturally, if the data are believed to be temporally dependent, then that information could be used to improve prediction.  However, developing a sound model for dependent data can be a challenge and model-based inference/prediction could be severely biased when based on a misspecified model.  Ideally, the data analyst could work with a relatively simple model for temporally dependent data and, if necessary, the data would suggest when some adjustments might help to accommodate model misspecification.  This is what the GPrC algorithm aims to provide.  

For the process $Y_1,Y_2,\ldots$, let $P^\star$ denote the true distribution and $P_\theta$ a posited model; note the slight abuse of notation letting, e.g., $P^\star$, denote here the full joint distribution whereas $P^\star$ stood for the marginal distribution of an individual $Y_i$ in the previous sections.  For the model $P_\theta$, we will be considering a simple, Gaussian, first-order autoregressive process that posits 
\[ (Y_{i+1} \mid Y_i=y') \sim \nm(\rho y', \sigma^2), \quad i=0,1,\ldots, \]
where $\theta=(\rho,\sigma) \in (-1,1) \times (0,\infty)$.  Of course, other more sophisticated models are possible; we opt for a simple and concrete model here to showcase the GPrC algorithm's ability to overcome model misspecification biases.  

As before, given the posited model and the observed data $Y^n=(Y_1,\ldots,Y_n)$, one can carry out a Bayesian analysis that leads to a posterior distribution, $\Pi_n$, for $\theta$.  Given that the model assumes a Markov or one-step temporal dependence, the $\eta$-generalized predictive distribution in this case has a density of the form 
\[ f_n^{(\eta)}(y \mid y') \propto \int p_\theta^\eta(y \mid y') \, \Pi_n(d\theta), \]
where $p_\theta(\cdot \mid y')$ is the $\nm(\rho y', \sigma^2)$ density.  Of course, this predictive density has quantiles, and we denote the upper-$\alpha$ quantile by 
\[ Q_\alpha(\eta; Y_n, Y^n), \quad \alpha \in (0,1). \]
Note here the dependence on $Y_n$ in two places: first, as a component in the data $Y^n$ and, second, in that the predictive density for $Y_{n+1}$ depends explicitly on $Y_n$.  

For the GPrC algorithm, all that is left to specify is the coverage probability and a bootstrap approximation.  Of course, the coverage probability function is 
\[ c_\alpha(\eta) = P^\star\{ Q_\alpha(\eta; Y_n, Y^n) \geq Y_{n+1}\}, \]
where the probability is with respect to the joint distribution of $Y^{n+1}$ determined by $P^\star$.  For a bootstrap approximation, it is important that the particular choice of bootstrap respects the temporal dependence.  To achieve this, we apply the block bootstrap strategy proposed by \citet{kunsch1989jackknife}; see, also, \citet{politis1994stationary} and  \citet[][Ch.~8]{davison1997bootstrap}. The basic idea behind the block boostrap is, as the name suggests, to resample blocks of observations with the goal of retaining the temporal dependence structure.   The version we employ here in our illustration is as follows. Select a fix block length parameter $\ell < n$ such that the number of blocks of a time series,  $k=n/\ell$, is an integer. For each bootstrap sample, we construct the bootstrap sample $Y_b^n$ by concatenating the results of sampling $k$ overlapping blocks of subsequences
\[ \{(Y_{s_j+1},\ldots, Y_{s_j+\ell}): j=1,\ldots,k\}, \]
where $\{s_1,\dots,s_k\}$ are the starting points of each block, and are generated from a discrete uniform distribution on $\{1,2,\dots,n-l+1\}$. Here we set the block length $\ell \sim n^{1/3}$ according to the recommendations in \citet{buhlmann2002bootstraps}; see, also \citet{gotze1996second} and \citet{buhlmann1999block}.

Given the block bootstrap samples $Y_b^n$ for $b=1,\ldots,B$, the empirical coverage probability is approximated as 
\begin{equation}
\label{eq:cvg.time}
\hat c_\alpha(\eta) = \frac1B \sum_{b=1}^B \Bigl[ \frac{1}{n-1} \sum_{i=1}^{n-1} 1\{Q_\alpha(\eta; Y_i, Y_b^n) \geq Y_{i+1}\} \Bigr].
\end{equation}
Note that the inner sum in \eqref{eq:cvg.time} ranges over $i=1,\ldots,n-1$, covering all the consecutive pairs $(Y_i,Y_{i+1})$ in the data, so $Y_n$ is never used as a direct argument in the $Q_\alpha(\eta; \cdot, Y_b^n)$ function. However, after $\hat\eta$ is determined and it is time to predict $Y_{n+1}$, we would take $Q_\alpha(\hat\eta; Y_n, Y^n)$ as our $100(1-\alpha)$\% upper limit, with $Y_n$ plugged in.  More generally, if the model posited lag-$m$ dependence, then the inner sum in \eqref{eq:cvg.time} would range over $i=1,\ldots,n-m$, covering all consecutive $m$-tuples $(Y_i,\ldots,Y_{i+m-1})$.  

Next, we investigate the performance of the GPrC-modified predictive distribution in in three different simulation scenarios:
\begin{enumerate}
\item First-order autoregressive with Laplace errors, i.e., $Y_{i+1} = 0.9Y_i + \eps_i$, for $i=1,\ldots,n-1$, where the $\eps_i$'s are iid $\lap(0,1)$; 

\vspace{-2mm}

\item Nonlinear time series with Laplace errors, i.e., $Y_{i+1} = \sin(Y_i) + \eps_i$, for $i=1,\ldots,n-1$, where the $\eps_i$'s are iid $\lap(0,1)$; 

\vspace{-2mm}

\item Nonlinear time series with heteroscedastic Laplace errors, i.e., $Y_{i+1} = \sin(Y_i) + (0.5+0.25Y_i^2)^{1/2}\eps_i$, for $i=1,\ldots,n-1$, where the $\eps_i$'s are iid $\lap(0,1)$; 

\end{enumerate}


The first scenario is one where the temporal dependence structure is correctly specified but the error distribution has heavier-than-normal tails. The second is one where both the temporal dependence structure and the error distribution of the posited model are misspecified. Finally, the third scenario is one where the temporal dependence structure of the posited model is misspecified, and the distribution of the errors is heteroskedastic with heavy-tailed distribution. The choice of ``0.90'' in the first scenario ensures that there is relatively strong temporal dependence in the true data-generating process. 

For comparison, we consider two standard methods.  The first method, which we call the {\em plug-in} method, is quite basic and is a natural choice when the auto-regressive model is assumed to be correct.  That is, the plug-in method produces maximum likelihood estimates, $\hat\theta = (\hat\rho, \hat\sigma)$, for the model parameter $\theta=(\rho,\sigma)$ and returns the $100(1-\alpha)$\% prediction limit for $Y_{n+1}$ as $\hat\rho Y_n + \Phi^{-1}(1-\alpha) \, \hat\sigma$.  Of course, if the model is correctly specified, then this would be approximately calibrated.  The second method is the proposed GPrC algorithm, that starts with a simple Bayesian model and then uses the data to tune $\eta$ as described above. In particular, we take a conjugate normal--inverse gamma to get the Bayesian posterior.  
Table~\ref{table:time.series} summarizes the coverage probability for the two different methods, with $n \in \{100,200,400\}$ at level $\alpha=0.01$. The plug-in method under covers at this extreme $\alpha$ level, whereas the GPrC method is close to the target coverage even with a sample size as low as $n = 100$.


\begin{table}[t]
\begin{center}
\begin{tabular}{ccccc}
 \hline
Model  & Method  & $n=100$ & $n=200$& $n=400$ \\
 \hline
1 & GPrC & 0.992 & 0.994 & 0.994 \\
& Plug-in & 0.976 & 0.973 & 0.981\\
\hline
2 & GPrC & 0.994 & 0.996 & 0.992 \\
& Plug-in & 0.978 & 0.976 & 0.975 \\
\hline
3 &GPrC & 0.990 & 0.990 & 0.992 \\
&Plug-in & 0.978 & 0.977 & 0.980 \\
\hline
\end{tabular}
\end{center}
\caption{Empirical coverage probability for the two methods' predictive distribution quantiles, at level $\alpha =0.01$, for the three {\em time series} model scenarios described in the text, based on 1000 replications.}
\label{table:time.series}
\end{table}

\subsection{Spatial data}
\label{S:result.spatial}

Let $\S \subseteq \RR^2$ denote a spatial region on which a stochastic process $Y = \{Y(s): s \in \S\}$ is defined.  For example, $\S$ may denote a set of geographical locations (expressed in some coordinate system) and $Y(s)$ denotes the temperature, precipitation level, etc.~at location $s \in \S$.  Data consists of a finite collection $s^n = (s_1,\ldots,s_n)$ of locations at which observations are made, along with the vector of $Y$-process measurements
\[ Y(s^n) = \bigl( Y(s_1), \ldots, Y(s_n) \bigr). \]
The goal is to predict the value of $Y(s_{n+1})$ at a new spatial location $s_{n+1} \in \S$.  

Model-based predictions are common in spatial applications \citep[e.g.,][]{stein2012interpolation}.  A model $P_\theta$ would consist of certain assumptions about the distribution of the stochastic process $Y$.  A common choice is to model $Y$ as a Gaussian process and the parameter $\theta$ characterizes its mean and covariance functions.  To make this characterization relatively simple, it is tempting to make rather strong assumptions, e.g., stationarity and/or isotropy, in addition to Gaussianity.  Such assumptions can be hard to justify, so working with such a simple model opens the data analyst up to risk of model misspecification bias.  Therefore, it would be interesting to see if a GPrC adjustment on top of a simple model could calibrate predictions and alleviate some of the data analyst's risk.  

In particular, we consider a simple Gaussian process model \citep[e.g.,][]{finley2019efficient} for $Y$, generically denoted by $P_\theta$, that posits a constant mean function $E\{Y(s)\} \equiv \mu$ an exponential covariance function 
\begin{equation}
\label{eq:cov.fun}
\C_\theta(s,t) = \sigma^2 \bigl( e^{-\|s-t\|/\hat\phi} + \hat\tau 1\{s=t\} \bigr), \quad s,t \in \S, 
\end{equation}
where $\theta=(\mu, \sigma^2)$ is treated as the unknown model parameter, while (variogram-based) plug-in estimates $\hat\phi$ and $\hat\tau$ of the range and scaled nugget parameters are treated as fixed and known.  (Some applications might have covariates that could be incorporated into the model's mean function, but we will not consider this here.)  Different from our previous sections with little or no dependence built into the model, here the model itself involves non-negligible dependence.  So the in-model conditional density function for $Y(s_{n+1})$, given $s_{n+1}$ and data $D^n=\{s^n, Y(s^n)\}$, is normal with  
\begin{align*}
m_\theta(s_{n+1}) & = \mu + \gamma_\theta(s^{n+1})^\top \Gamma_\theta^{-1}(s^n) \{Y(s^n) - \mu 1_n\} \\
v_\theta(s_{n+1}) & = C_\theta(s_{n+1}, s_{n+1}) - \gamma_\theta(s^{n+1})^\top \Gamma_\theta^{-1}(s^n) \gamma_\theta(s^{n+1}),
\end{align*}
where $1_n$ is an $n$-vector of unity, $\gamma_\theta(s^{n+1})$ is an $n$-vector with $i^\text{th}$ entry $C_\theta(s_i, s_{n+1})$, and $\Gamma_\theta(s^n)$ is an $n \times n$ matrix with $(i,j)^\text{th}$ entry $C_\theta(s_i, s_j)$.  Obviously, the density function of the aforementioned conditional distribution depends on the unknown $\theta$, so we denote this by $p_\theta(\cdot \mid s_{n+1}, D^n)$.  We complete the Bayesian model formulation by introducing a conjugate normal--inverse gamma prior for $\theta$, from which we immediately obtain a posterior distribution $\Pi_n$ for $\theta$ depending on data $D^n$.  Then we define our $\eta$-generalized predictive density for $Y(s_{n+1})$, given $s_{n+1}$ and data $D^n$, as
\[ f_n^{(\eta)}(y \mid s_{n+1}) \propto \int p_\theta^\eta(y \mid s_{n+1}, D^n) \, \Pi_n(d\theta). \]
The GPrC method aims to use information in the available data to tune the learning rate $\eta$ in order to calibrate this predictive distribution in the event that the simple conjugate Gaussian process model is misspecified.

Of course, the GPrC algorithm relies on the bootstrap.  Since these spatial data models are more complex than those in previous sections, naturally an appropriate bootstrap procedure will be similarly more complex.  The procedure we consider here is the so-called {\em semi-parametric bootstrap} procedure developed in \citet{schelin2010kriging}.  This procedure is ``semi-parametric'' in the sense that it assumes the model's mean and covariance functions are correctly specified, but does not rely on any distributional assumptions, such as Gaussianity.  Start by obtaining suitable estimates $\hat\theta$ of the model parameter $\theta$; the scale $\sigma^2$ is typically estimated through a sample variogram, while $\mu$ can be estimated using a simple average of the observed $Y(s^n)$'s. Next, define the $n$-vector of residuals 
\[ \eps(s^n) = \Gamma_{\hat\theta}^{-1/2}(s^n) \{Y(s^n) - \hat\mu 1_n\}. \]
The idea is that the residuals are roughly free of any spatial dependence, i.e., that they are approximately iid.  Let $B$ be the desired number of bootstrap samples.  Now, for $b=1,\ldots,B$, take a sample of {\em size $n+1$}, with replacement from the $n$-vector of residuals and denote these as 
\[ \eps_b(s^{n+1}) = \bigl( \eps_b(s_1), \ldots, \eps_b(s_n), \eps_b(s_{n+1}) \bigr)^\top, \quad b=1,\ldots,B. \]
Then we just add back the mean and spatial dependence in the natural way, 
\[ Y_b(s^{n+1}) = \hat\mu 1_n + \Gamma_{\hat\theta}^{1/2}(s^{n+1}) \, \eps_b(s^{n+1}), \quad b=1,\ldots,B, \]
where 
\[ \Gamma_\theta(s^{n+1}) = \begin{pmatrix} \Gamma_\theta(s^n) & \gamma_\theta(s^{n+1}) \\ \gamma_\theta(s^{n+1})^\top & C_{\hat\theta}(s_{n+1}, s_{n+1}) 
\end{pmatrix} \] 
Finally, at a candidate $\eta$, and for each $b=1,\ldots,B$, the GPrC will use the first $n$ components, $Y_b(s^n)$, of $Y_b(s^{n+1})$, to construct the generalized predictive distribution and extract the quantile $Q_\alpha(\eta; s_{n+1}, Y_b(s^n))$.  And then the coverage probability is approximated as 
\[ \hat c_\alpha(\eta) = \frac1B \sum_{b=1}^B 1\{ Q_\alpha(\eta; s_{n+1}, Y_b(s^n)) \ni Y_b(s_{n+1})\}, \]
where, note, the predictive distribution quantile is tested against the corresponding last entry, $Y_b(s_{n+1})$, of $Y_b(s^{n+1})$.  This process is repeated on the sequence of $\eta$ values as defined the GPrC algorithm until convergence to some $\hat\eta$.  

Other versions of the spatial bootstrap are possible.  For example, if the model's mean function were non-constant, say a parametric function of the spatial coordinates and/or covariates, then this could be readily accommodated by adjusting how the residuals above are calculated.  Different variations on the bootstrap procedure itself are also available \citep[e.g.,][]{castillo2019nonparametric} whose use in the GPrC algorithm will be explored in subsequent work.  



To investigate the performance of our proposed GPrC for prediction in this spatial data context, we consider three different data-generating processes, as described below; the first two scenarios were investigated in \citet{schelin2010kriging}, while the third is from \citet{sang2010continuous}.  For each scenario, the $n$ spatial locations $s^n$ at which the $Y$ process is measured are uniformly distributed in a disc of radius $r = 20$ around $(0, 0)$, and the target location is set at $s_{n+1} = (0, 0)$.

\begin{enumerate}
\item A (correctly specified) Gaussian process model as described above, with $\mu^\star=0$, $\sigma^\star=1$, $\phi^\star=3$, and $\tau^\star=0$.  

\vspace{-2mm}

\item Following \citet{schelin2010kriging}, let $L_{n+1}$ denote a vector of iid $\logn(0,1)$ random variables and define the response $Y$ as  
\[ Y(s^{n+1}) =  \log\Bigl[ \Gamma_\theta^{1/2} \bigl\{ L_{n+1} / (e^2-e^1)^{1/2} \bigr\} \Bigr], \]
where $\mu^\star=0$, $\sigma^\star=1$, $\phi^\star=3$, and $\tau^\star=0$, with log taken component-wise.  

\vspace{-2mm}

\item Following \citet{sang2010continuous}, let the response $Y$ be defined through the following hierarchical generalized extreme value process.  Start with two independent Gaussian processes, say, $Z_1$ and $Z_2$, with covariance functions having the same exponential form as in \eqref{eq:cov.fun}, with associated covariance parameters $(\mu_1, \sigma_1, \phi_1, \tau_1) = (10,1,4,0)$ and $(\mu_2, \sigma_2, \phi_2, \tau_2) = (0,1, 1.4, 0)$.  Let $\Phi$ and $G$ denote the distribution functions of the standard normal and standard Fr\'echet distributions, respectively, and then the define the process $W$ as 
\[ W(s) = Z_1(s) + \sigma \xi^{-1} \bigl( [G^{-1} \circ \Phi\{Z_2(s)\}]^\xi - 1 \bigr), \]
where $(\sigma, \xi) = (3, 0.5)$.  Like the log-normal setup in Scenario~2 above, since the mean $\mu_1$ of $Z_1$ is rather large, it is virtually impossible for $W(\cdot)$ to be negative.  Therefore, we take the response process as $Y(s) = \log W(s)$.
\end{enumerate}

We compare the performance of GPrC with that of two alternative approaches.  One is a simple {\em plug-in} approach that starts with variogram-based estimates $\hat\theta$ of the basic model parameters and constructs an upper prediction limit for the response at a new location $s_{n+1}$ as 
\[ \widehat Q_\alpha(D^n) = m_{\hat\theta}(s_{n+1}) + \Phi^{-1}(1-\alpha) \, \{ v_{\hat\theta}(s_{n+1}) \}^{1/2}. \]
Of course, if the process is Gaussian, then this would be an approximately valid $100(1-\alpha)$\% prediction upper limit.  However, if the proposed Gaussian process model is misspecified, the plugin predictive interval can result in a narrower predictive interval.  Second, we consider a more robust semi-parametric {\em bootstrap} method based directly on the output from the bootstrap procedure in \citet{schelin2010kriging}.  That is, we let $\widehat Q_\alpha(D^n)$ be the upper-$\alpha$ quantile of the empirical distribution of $\{Y_b(s_{n+1}): b=1,\ldots,B\}$.  The three prediction limits are compared based on coverage probability and relative interval scores as described in Section~\ref{SS:sims}.  The results are presented in Table~\ref{table:spatial}.





\begin{table}[t]
\begin{center}
\begin{tabular}{cccccccc}
    \multicolumn{2}{c}{} &  \multicolumn{3}{c}{Coverage Probability} & \multicolumn{3}{c}{Interval score}\\
 \toprule
Scenario & Method & $90$\% & $95$\% & $99\%$ & $90$\% & $95\%$& $99\%$\\
 \toprule
1 & GPrC & 0.888 & 0.940  & 0.994 & 1.718 & 2.062 & 2.501\\
 & Bootstrap & 0.882 & 0.932 & 0.988 & 1.715 & 2.072 & 2.549\\
 & Plug-in & 0.889 & 0.946 & 0.993 & 1.707 & 2.022 & 2.420\\
  \hline
2 & GPrC & 0.896 & 0.957 & 0.985 & 1.288 & 1.476 & 2.148 \\
 & Bootstrap & 0.896 & 0.952 & 0.985 & 1.293 & 1.485 & 2.154\\
 & Plug-in  & 0.862 & 0.933 & 0.974 & 1.278 & 1.394 & 2.220 \\
  \hline
3 & GPrC & 0.888 & 0.941 & 0.988 & 3.636 & 3.857 & 4.737\\
 & Bootstrap & 0.880 & 0.931 & 0.984 & 3.633 & 3.878 &  4.977\\
 & Plug-in & 0.865 & 0.902 & 0.950 & 3.647 & 4.002 & 5.765\\
\bottomrule
\end{tabular}
\end{center}
\caption{Empirical coverage probabilities and relative interval scores based on $1000$ data sets of size $n=100$, for each of the three scenarios, in the {\em spatial data} example.}
\label{table:spatial}
\end{table}

In Scenario~1, the correctly specified Gaussian process, all three methods perform well in terms of coverage and interval score, as expected.  Perhaps GPrC has a slight advantage over bootstrap in terms of interval score at the extreme 99\% quantile, but both are comparable to plug-in method which would be (at least close to) optimal in this correctly specified model setting.  In Scenarios~2 and 3, both where the Gaussian process model is misspecified, the plug-in method suffers in terms of under-coverage, as expected.  Surprisingly, both GPrC and the bootstrap are able to overcome the model misspecification and reach nearly the target coverage across the board, while maintaining some amount of efficiency as indicated by the interval scores.  In particular, although GPrC is making use of the semi-parametric bootstrap to approximate coverage probabilities, its predictions are still effectively model-based.  So the fact that calibration can be (approximately) achieved using a model-based Bayesian method in a challenging application under fairly severe model misspecification is remarkable. 



\section{Conclusion}
\label{S:discuss}

Following up on recent work that tunes the learning rate parameter in generalized posterior distributions, in this paper we developed a procedure to calibrate generalized predictive distributions.  The idea is the upper-$\alpha$ quantiles of one's (subjective) predictive distribution ought to have an alternative (objective) interpretation as a valid, $100(1-\alpha)$\% prediction upper limit.  Our proposal---the GPrC algorithm---is to construct a generalized predictive distribution that depends on a tuning parameter $\eta$, approximate the coverage probability of its $\eta$-dependent prediction upper limit, and then tune $\eta$ in order to match the target coverage.  The key step is the coverage probability approximation, which we do via bootstrap, and we demonstrated numerically that calibration can be achieved via the proposed GPrC algorithm in a variety of settings using appropriate bootstrap procedures. 

In Section~\ref{SS:more} we presented a combination of heuristic and numerical arguments to support the claim that GPrC does, indeed, achieve calibration, at least approximately.  The challenge to providing a rigorous mathematical proof of this conjecture is that the {\em combination} of at least two powerful computational tools---bootstrap, stochastic approximation, and sometimes Monte Carlo sampling---each theoretically sound on its own, adds a level of complexity that makes the GPrC algorithm's dynamics very difficult to analyze.  The same is true for the GPC algorithm of \citet{syring2019calibrating}, so further theoretical understanding of how these methods work is an interesting open problem. 

The focus of this paper was on developing a general method for calibrating Bayesian-like predictive distributions under model misspecification and demonstrating that this method can be used in a relatively wide range of applications.  It would be interesting to investigate a particular application, e.g., in the spatial domain, to tailor GPrC to that specific application, and push the limits of how complex the models can be while still achieving approximate calibration.

\section*{Acknowledgments}

The authors thank Nicholas Syring for helpful comments on an initial draft.  This work is partially supported by the U.S.~National Science Foundation, grants DMS--1811802 and SES--2051225.

\bibliography{gprc}

\end{document}